\documentclass[10pt,journal]{IEEEtran}

\ifCLASSINFOpdf
\else
\fi
\usepackage[cmex10]{amsmath}
\usepackage{mathrsfs}
\usepackage{graphicx}
\usepackage{float}
\usepackage{amsfonts,amssymb}
 \usepackage{color}
 \usepackage{subfigure}
 \usepackage{booktabs}
 \usepackage{amsthm}
\usepackage{subfig}
\usepackage{multirow}
\usepackage{xcolor}
\usepackage{cite}

\usepackage{graphicx}
\usepackage{epstopdf}

\theoremstyle{plain} \newtheorem{property}{Property}
\theoremstyle{plain} \newtheorem{proposition}{Proposition}
\theoremstyle{plain} \newtheorem{algorithm}{Algorithm}

\begin{document}
%
\title{JPEG Noises beyond the First Compression Cycle }

\author{Bin Li,
        Tian-Tsong Ng,
        Xiaolong Li,
        Shunquan Tan,
        and Jiwu Huang
\thanks{Bin Li is with the Shenzhen Key Lab of Modern Communications and Information Processing, Shenzhen University, Shenzhen 518060, China
(phone: 86-755-2267-3509; fax: 86-755-2653-6122; e-mail: libin@szu.edu.cn).}
\thanks{Tian-Tsong Ng is with Institute for Infocomm Research, A*STAR, 138632 Singapore (e-mail: ttng@i2r.a-star.edu.sg).}
\thanks{Xiaolong Li is with the Institute of Computer Science and Technology, Peking University, Beijing 100871, China (e-mail: lixiaolong@pku.edu.cn).}
\thanks{Shunquan Tan is with School of Computer Science and Software Engineering, Shenzhen University, Shenzhen 518060, China (email: tansq@szu.edu.cn).}
\thanks{Jiwu Huang is with the College of Information Engineering, Shenzhen University, Shenzhen 518060, China, (e-mail: jwhuang@szu.edu.cn).}
}


\maketitle

\begin{abstract}
This paper focuses on the JPEG noises, which include the quantization noise and the rounding noise, during a JPEG compression cycle.
The JPEG noises in the first compression cycle have been well studied; however, so far less attention has been paid on the JPEG noises in higher compression cycles.
In this work, we present a statistical analysis on JPEG noises beyond the first compression cycle.
To our knowledge, this is the first work on this topic.
We find that the noise distributions in higher compression cycles are different from those in the first compression cycle, and they are dependent on the quantization parameters used between two successive cycles.
To demonstrate the benefits from the statistical analysis,
we provide two applications that can employ the derived noise distributions to uncover JPEG compression history with state-of-the-art performance.
\end{abstract}


\IEEEpeerreviewmaketitle


\section{Introduction}

\IEEEPARstart{L}{ossy} JPEG compression \cite{Wallace1991} achieves a good compression ratio but the process consists of two main irreversible steps that result in image quality degradation. The first lossy step is the quantization of the $8 \times 8$ block-DCT coefficients. The quantization step size for each block-DCT coefficient is specified in a quantization table; the larger the step size leads to a larger loss of image information. The second lossy step corresponds to the integer rounding operation when JPEG coefficients are restored into image pixel representation. An image can be JPEG-compressed multiple times and the image quality may suffer increasing degradation at each cycle of lossy JPEG compression.

The changes in image signal due to the mentioned lossy steps are often considered as noise. We can refer to the signal loss due to the first kind of lossy step as \emph{quantization noise}, and the second kind as \emph{rounding noise}.
We refer to them as \emph{JPEG noises}.
The JPEG noises have been studied in the past for a single JPEG compression scenario \cite{Yovanof1996,Robertson2005,Luo2010}. Such studies may be useful for JPEG image quality restoration \cite{Chou1998, Robertson2005}, JPEG transcoding \cite{Coulombe2010}, and JPEG compression history identification \cite{Fan2003,Neelamani2006, Chen2009,Pevny2008,Farid2006,Luo2010,Huang2010,Qu2008,Lin2011}.

To our knowledge, no studies of JPEG noises has explored beyond the first JPEG compression cycle. The main reason could be due to the common assumption that the properties of JPEG noises of a higher compression cycle is similar to that of the first cycle. The second reason could be due to the lack of applications that require the knowledge of JPEG noises of the second compression cycle and beyond.

In this paper, we derived the statistical properties of JPEG noises for higher compression cycles.
An interesting result indicates that the distributions of JPEG noises in a higher compression cycle are not always the same as those of the first cycle.
Apart from introducing JPEG noise distributions, we also introduced various bounds on their distribution parameters that concerns with natural images. Such analysis is useful in two application scenarios, \emph{i.e.} JPEG quantization step estimation  \cite{Fridrich2001,Fan2003,Neelamani2006,Ye2007,Luo2010,Lin2011}, and JPEG re-compression identification \cite{Huang2010}, and leads to superior performance.

The rest of this paper is organized as follows.
We provide the preliminaries on JPEG compression and JPEG noises in the next section.
The bounds of noise variances in the first JPEG compression cycle are summarized in Section III to facilitate our discussion on statistical properties of JPEG noises during multiple JPEG cycles.
In Section IV, we present statistical analysis on JPEG noises in higher compression cycles.
Two applications of the derived theoretical models are given in Section V.
We conclude our paper in Section VI. The proof for the new propositions are given in respectively Appendixes.

\begin{figure}[!tpb]
    \centering
    \includegraphics[width=0.45\textwidth]{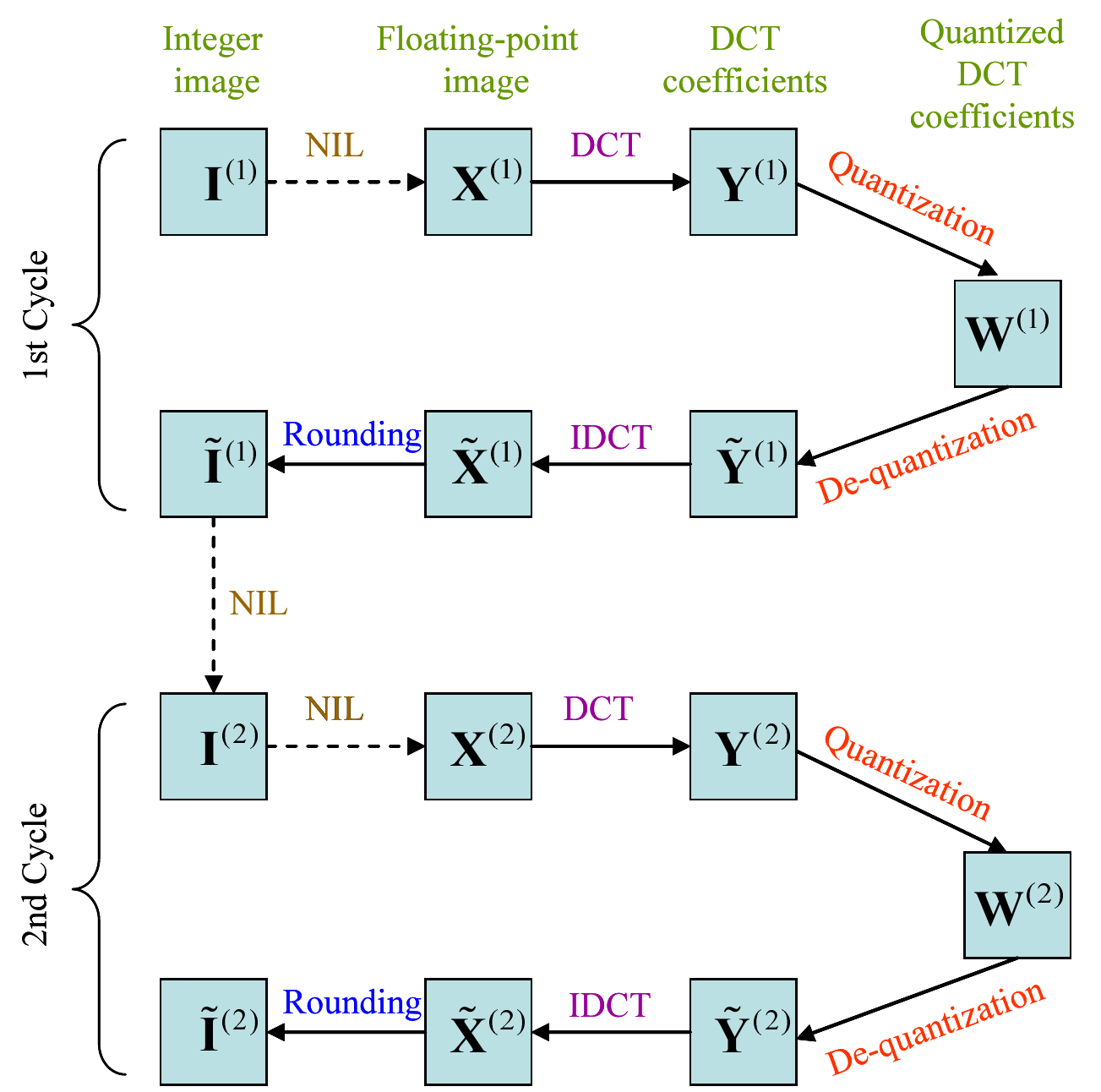}
    \caption{Processing steps for multi-cycle JPEG compression. The NIL symbols refer to no operations.}
    \label{fig:flowchart1}
\end{figure}

\section{Preliminaries}

In this section, we present some of the common knowledge about JPEG compression and JPEG noises. The results presented in this section are either can be found in the prior work \cite{Yovanof1996,Fan2003,Robertson2005,Luo2010} or can be derived easily based on the prior work.

\subsection{JPEG Compression}

In JPEG compression, an image undergoes a series of processing and transformation. The operations are denoted by the directed arrows in Fig.~\ref{fig:flowchart1} where the NIL symbols that denote no operations are meant for preserving the symmetry in the diagram. Each JPEG compression cycle is composed of an encoding phase and a decoding phase.
In an encoding phase, an image is converted from an integer representation to a floating-point representation, before being transformed into $8 \times 8$ block-DCT coefficients.
The DCT coefficients are quantized into integers by specific quantization steps.
The quantized integers will be losslessly encoded and stored in JPEG file format.
The decoding phase is essentially the reverse of the encoding phase. Information loss happens during the quantization step at the encoding phase and the integer rounding step at the end of the decoding phase.

\begin{figure}[!tpb]
    \centering
    \includegraphics[width=0.40\textwidth]{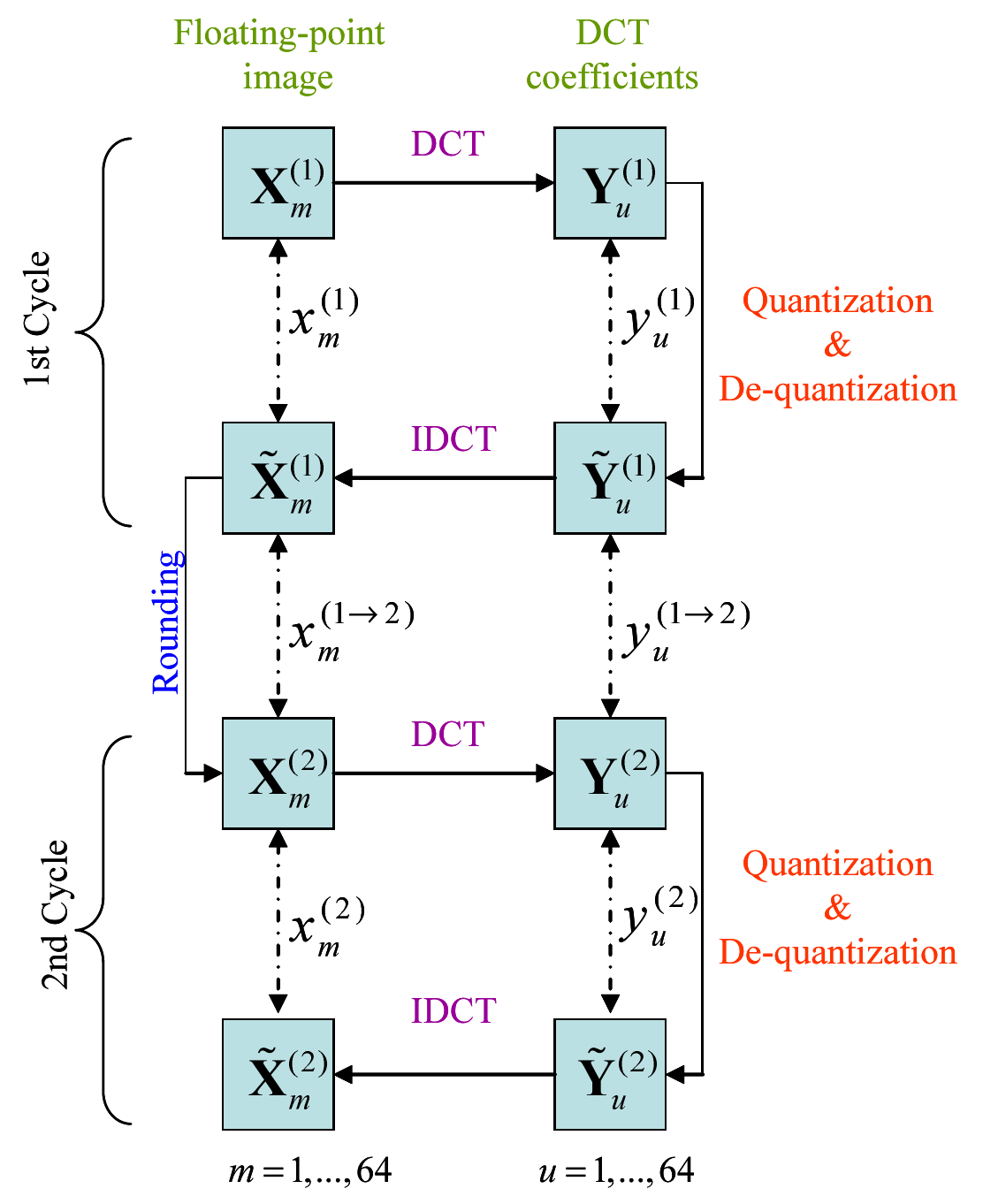}
    \caption{Logical diagram for multi-cycle JPEG compression}
    \label{fig:flowchart2}
\end{figure}

\subsection{JPEG Noises}

By ignoring the NIL operations in Fig.~\ref{fig:flowchart1}, we can obtain a simpler logical diagram for multi-cycle JPEG compression as shown in Fig.~\ref{fig:flowchart2}, in which the lossy operations are shown more prominently. Fig.~\ref{fig:flowchart2} also shows the JPEG noises where $y^{(\cdot)}$ in the DCT domain are the quantization noises and $x^{(\cdot \rightarrow \cdot)}$ in the image pixel domain are the rounding noises. Note that the rounding noises are considered inter-compression-cycle noises in our setting.

In this work, we assume no shifting of DCT block structure between successive compression cycles. Due to the DCT $8 \times 8$ block structure, the image pixels and DCT coefficients can be respectively indexed by $m$ and $u$ from 1 to 64. Henceforth, whenever there is no confusion of per-pixel or per-DCT-coefficient operations, we will not indicate the index from the notation.

\vspace{0.3cm}
\noindent \textbf{Quantization noise:} Quantization noise is defined as
\begin{equation}\label{eq:Nq1}
    y^{(k)} =  Y^{(k)} - \tilde{Y}^{(k)} = Y^{(k)} -  \left[  \frac{ Y^{(k)} }{q^{(k)}}  \right] q^{(k)}, \quad q^{(k)} \in \mathbb{N},
\end{equation}
where $Y^{(k)}$ and $\tilde{Y}^{(k)}$ are respectively DCT coefficients before quantization and after de-quantization, $q^{(k)}$ denotes the quantization step size, and $[\cdot]$ represents the integer rounding operation.

\vspace{0.3cm}
\noindent \textbf{Rounding noise:} Rounding noise is simply defined as
\begin{equation}\label{eq:Nr}
    x^{(k \rightarrow k+1)} =  \tilde{X}^{(k)} - X^{(k+1)},
\end{equation}
where $\tilde{X}^{(k)}$ and $X^{(k+1)}$ correspond to spatial-domain images of two adjacent compression cycles, where the former is in floating point and the latter is in integer. Note that, rounding operation is essentially a special quantization operation with $q^{(k)} = 1$, where it can be rewritten as:
\begin{equation}\label{eq:Nr1}
    x^{(k \rightarrow k+1)} =  \tilde{X}^{(k)} - [\tilde{X}^{(k)}].
\end{equation}

\vspace{0.3cm}
\noindent \textbf{Auxiliary noises:} Apart from the quantization noise and rounding noise, two other type of noises, \emph{i.e.}, $x^{(\cdot)}$ in the image pixel domain and $y^{(\cdot \rightarrow \cdot)}$ in the DCT domain, are introduced as a side-product to facilitate mathematical derivation.

The \emph{spatial auxiliary noise} is defined as:

\begin{equation}\label{eq:Nx}
    x^{(k)} =  X^{(k)} - \tilde{X}^{(k)},
\end{equation}
where both $X^{(k)}$ and $\tilde{X}^{(k)}$ belong to the same compression cycle.
The \emph{DCT auxiliary noise} is defined as:

\begin{equation}\label{eq:Ny}
    y^{(k \rightarrow k+1)} =  \tilde{Y}^{(k)} - Y^{(k+1)},
\end{equation}
where $\tilde{Y}^{(k)}$ and $Y^{(k+1)}$ belong to different compression cycles.

\vspace{0.3cm}
\noindent \textbf{Relationship between noises:} The noises are interconnected. From Equation~(\ref{eq:Nx}), we can derive:
\begin{equation}\label{eq:Nx1}
    \begin{split}
        x^{(k)}
        &=  \textit{IDCT}(Y^{(k)}) - \textit{IDCT}(\tilde{Y}^{(k)}) \\
        &=  \textit{IDCT}(Y^{(k)} - \tilde{Y}^{(k)}) = \textit{IDCT}(y^{(k)}),
    \end{split}
\end{equation}
where $x^{(k)}$ is simply the inverse DCT transform of the quantization noise $y^{(k)}$. The derivation above is based on the linear property of the IDCT operator.

Similarly, from Equation~(\ref{eq:Ny}), we obtain:
\begin{equation}\label{eq:Ny1}
\begin{array} {lcl} y^{(k \rightarrow k+1)} & = & \textit{DCT}(\tilde{X}^{(k)}) - \textit{DCT}(X^{(k+1)}) \\
& = & \textit{DCT}(\tilde{X}^{(k)} - X^{(k+1)}) \\
& = & \textit{DCT}(x^{(k \rightarrow k+1)}), \end{array}
\end{equation}
where  \textit{DCT} represents the linear DCT operator.

Furthermore, from \eqref{eq:Nr1} and \eqref{eq:Nx}, we know that
\begin{equation}\label{eq:round_aux_relation}
\begin{split}
x^{(k \rightarrow k+1)} &= {\tilde{X}}^{(k)}    -  [  {\tilde{X}}^{(k)} ]  \\
&= {X}^{(k)} - {x}^{(k)} - [ {X}^{(k)} - {x}^{(k)} ] \\
&=  -  {x}^{(k)} + \left( {X}^{(k)} - [ {X}^{(k)} - {x}^{(k)} ] \right) \\
&= - ( {x}^{(k)} - [{x}^{(k)} ])
\end{split}
\end{equation}
Note that the last expression holds because both $  {X}^{(k)} $ and $  [  {X}^{(k)} -  {x}^{(k)} ]$ are integers.
As a result, we can regard the rounding noise $x^{(k \rightarrow k+1)}$ as arising from negatively rounding off the auxiliary noise ${x}^{(k)}$,
or a kind of negative quantization noise of ${x}^{(k)}$ with quantization step size of 1.

\vspace{0.3cm}
\noindent \textbf{Auxiliary noise distribution:}
The central limit theorem (CLT) states that the (weighted) summation of many independent and identically distributed random variables are approximately Gaussian.
If the random variables are not identical in distribution but satisfy the Lindeberg's condition \cite{RobertB.Ash2000},
which only requires them to be independent and have finite mean and variance,
their sum also tends to be Gaussian distributed.
The CLT still holds if there are mild correlations among random variables, as implied in \cite{Lam2000}.
Considering the IDCT relation between $x^{(k)}$ and $y^{(k)}$,
and the DCT relation between $y^{(k \rightarrow k+1)}$ and $x^{(k \rightarrow k+1)}$,
no matter what distributions the quantization noise $y^{(k)}$ and the rounding noise $x^{(k \rightarrow k+1)}$ follow,
the auxiliary noises $x_{}^{(k)}$ and $y^{(k \rightarrow k+1)}$
should follow Gaussian distribution.
This result on auxiliary noise is regardless of the location within an $8 \times 8$ block.

\subsection{General Distribution of Quantization Noise}

In general, the distribution for quantization noise as defined in \eqref{eq:Nq1} is given by:
\begin{equation} \label{eq:ypdf}
\begin{split}
f_y(s)
    &= \sum_{k = -\infty}^{\infty}f_Y (kq+s),   \quad s \in [-\frac{q}{2}, \frac{q}{2} ), k \in \mathbb{Z},
 \end{split}
\end{equation}
where $f_y$ and $f_Y$ are respectively the distribution for $y$ and $Y$, and $q$ is the quantization step size. Since integer rounding is a kind of quantization with $q=1$, \eqref{eq:ypdf} also applies to the rounding noise.

We will call $f_y$ a \emph{quantized-Laplacian distribution} if $Y$ belongs to zero-mean Laplacian distribution $\mathcal{L}(0, \lambda)$. We denote the quantized-Laplacian distribution as $\mathcal{Q^L}(\lambda, q)$, where $q$ is the quantization step size. Likewise, $f_y$ is called a \emph{quantized-Gaussian distribution} and denoted by $\mathcal{Q^N}(\sigma^2, q)$ if $Y$ belongs to a zero-mean Gaussian distribution $\mathcal{N}(0, \sigma^2)$.
Note that both the quantized-Laplacian distribution and the quantized-Gaussian distribution have zero-mean.

\subsection{JPEG noises of First Compression Cycle}

The distribution of JPEG noises for the first compression cycle can be specialized from \eqref{eq:ypdf}, if the distribution of the coefficients are known. The coefficients of $8 \times 8$ block-DCT are indexed from 1 to 64. The first coefficient ($u =1$) is the mean of all pixel values in an $8 \times 8$ block and is called a DC coefficient due to its low-pass property. The other coefficients ($u = 2,\ldots,64$) are high-pass in nature and are called AC coefficients.

\vspace{0.3cm}
\noindent \textbf{Quantization noise:} It is widely accepted that the DC component of DCT coefficient of an uncompressed natural image has a Gaussian distribution and the AC components are zero-mean Laplacian distributed \cite{Reininger1983}:

\begin{property} \label{pro:uncompressedPDF}
The distribution of DCT coefficients is as follows:
\begin{equation}\label{eq:jpeg_dct_pdf}
Y_{u}^{(1)} \sim
\begin{cases}
\mathcal{N}(\mu_{Y_{1}^{(1)}}, \sigma^2_{Y_{1}^{(1)}} ), \quad  &u = 1  \\
\mathcal{L}(\mu_{Y_{u}^{(1)}}, \lambda_{Y_{u}^{(1)}} ),\quad &u \in \{2,3, \cdots, 64\}
\end{cases}
\end{equation}
where $\mathcal{N}$ and $\mathcal{L}$ denotes respectively a Gaussian and a Laplacian distribution. Note that, in this case, $ \lambda_{Y_{u}^{(1)}} = \sqrt{2 / \sigma_{Y_u^{(1)}}^{2}}$.
\end{property}
\noindent
Therefore, the quantization noise corresponding to the AC components is distributed according to a quantized-Laplacian distribution. In general, the support for the DC component is considered very large, in comparison with typical quantization step sizes. Therefore, for typical quantization step sizes, the corresponding quantization noise would be close to be uniformly distributed. This result can be summarized as below:

\begin{property} \label{pro:round1q}
The quantization noise of the first compression cycle has the following distributions:
\begin{equation}\label{eq:jpeg_dct_qnoise}
y_u^{(1)} \sim
\begin{cases}
 \mathcal{U}(-\frac{q_1^{(1)}}{2}, \frac{q_1^{(1)}}{2}), \quad &u = 1  \\
 \mathcal{Q^L}(\lambda_{Y_u^{(1)}},q_u^{(1)}), \quad &u \in \{2,3, \cdots, 64\},
\end{cases}
\end{equation}
where $u$ is the index of the block-DCT coefficients, $q_u^{(1)}$ is the quantization step size of index $u$ in the first compression cycle, and $\mathcal{U}$ represents a uniform distribution with indicated lower and upper supports.
\end{property}

\vspace{0.3cm}
\noindent \textbf{Rounding noise:} Rounding happens at the end of a compression cycle.
At the end of the first compression cycle, the auxiliary noise $x_{}^{(1)}$ within an $8 \times 8$ block is considered to follow a zero-mean Guassian distribution.
By applying the quantization relation from \eqref{eq:round_aux_relation} between the rounding noise and the auxiliary noise,
the distribution of rounding noise can be summarized as follows:

\begin{property} \label{pro:round1r}
For all spatial index $m$, the rounding noise of the first compression cycle has the following distributions:
\begin{equation}\label{eq:rnoise}
x_m^{(1 \rightarrow 2)} \sim \mathcal{Q^N}(\sigma^2_{x_m^{(1)}},1).
\end{equation}
Note that the distribution of $x_m^{(1 \rightarrow 2)}$ is related to the variance of the auxiliary noise $x_m^{(1)}$.
\end{property}

\section{Bounds for First-cycle JPEG noises}

As the first-cycle JPEG noises is the starting point for deriving the properties of higher-cycle JPEG noises, it is important to understand the numerical properties of the noise variances in terms of upper bounds and how tight are the bounds in relation to quantization step $q$.

\vspace{0.3cm}
\noindent \textbf{Quantization noise:} As the effect of quantization of a random variable is essentially about slicing its distribution, the resultant quantization noise distribution would be the sum of the slices. Therefore, it is understandable that a quantization noise distribution will approach uniform distribution as the quantization step is small when compared to the variance of the random variable.
We justify this conclusion in Lemma \ref{pro:lemma} of Appendix \ref{appendixC}.
From Property~\ref{pro:round1q}, we can derive the upper of quantization noise as below:

\begin{property} \label{pro:round1q_bound}
For all DCT coefficient index $u$, the upper bound of the quantization noise variance is given by:
\begin{equation}\label{eq:round1q_bound}
\sigma^2_{y_u^{(1)}}
\begin{cases}
 = \left(q_1^{(1)}\right)^2 / 12, \quad &u = 1  \\
 \le \left(q_u^{(1)}\right)^2 / 12, \quad &u \in \{2,3, \cdots, 64\},
\end{cases}
\end{equation}
where $q_u^{(1)}$ is the quantization step of the index $u$ in the first compression cycle.
\end{property}

\vspace{0.3cm}
\noindent \textbf{Rounding noise:} As rounding is a special case for quantization, we can similarly derive the following property for rounding noise.

\begin{property} \label{pro:round1r_bound}
For all spatial index $m$, the upper bound of rounding noise is given by:
\begin{equation}\label{eq:round1r_bound}
\sigma^2_{x_m^{(1 \rightarrow 2)}} \le 1/12.
\end{equation}
\end{property}

\vspace{0.3cm}
\noindent \textbf{Auxiliary noises:} From the previous section, we show that the auxiliary noises are related to the quantization noise and rounding noise through IDCT or DCT operation, as shown in \eqref{eq:Nx1} and \eqref{eq:Ny1}. Because of this relationship, we can show that:

\begin{property} \label{pro:round1aux_bound}
The upper bounds of auxiliary noises are given by:
\begin{equation}\label{eq:round1aux_bound1}
\sigma^2_{x_m^{(1)}} \le \max_u \left\{\sigma^2_{y_u^{(1)}} \big| u \in \{1,\cdots,64\}  \right\}, \forall m
\end{equation}
and
\begin{equation}\label{eq:round1aux_bound2}
\sigma^2_{y_u^{(1 \rightarrow 2)}} \le \max_m \left\{ \sigma^2_{x_m^{(1 \rightarrow 2)}}
 \big| m \in \{1,\cdots,64\} \right\}, \forall u.
\end{equation}

\end{property}

\section{JPEG noises beyond the First Compression Cycle} \label{sec:secondcycle}

In this section, we move the statistical analysis of JPEG compression beyond the first compression cycle. We show that statistical properties of JPEG noises for the second and the higher compression cycle are rather different from that of the first. Such result will force us to ask which compression cycle is it before statistical analysis on JPEG images is applied, in contrast to the current practice that disregards such important information. The proof of all propositions in this section can be found in Appendixes.

\subsection{JPEG noises of Second Compression Cycle} \label{ssec:s-cycle}

The analysis for the second compression cycle builds on the statistical properties for the first cycle, as stated in Property~\ref{pro:round1q} and Property~\ref{pro:round1r}. This represents a very different starting point as compared to that of the first compression cycle, hence, it is understandable that it leads to different statistical properties for the second cycle.

\vspace{0.3cm}
\noindent \textbf{Quantization noise:} We studied three cases for the second-cycle quantization noise. The first one is when $q_u^{(2)} = 1$. The second one is when $q_u^{(2)} \ge 2$ and when the second-cycle quantization step is an integer divisor of the first-cycle quantization step, \emph{i.e.}, $q_u^{(1)} = k_u q_u^{(2)}$, where $k_u$ is an integer. The third case covers the remaining possibilities. As the second case appear repeatedly in the text, we henceforth refer to it as the \emph{divisible quantization condition}.
Our analysis leads to the following proposition.

\begin{proposition} \label{pro:secondq} For all DCT coefficient $u$, the second-cycle quantization noise follows the following distributions:
\begin{equation}\label{eq:jpeg_dct_qnoise}
y_u^{(2)} \sim
\begin{cases}
 \mathcal{Q^N}( \sigma^2_{y_u^{(1 \rightarrow 2)}}   ,1), \quad &q_u^{(2)} = 1  \\
 \mathcal{N}( 0, \sigma^2_{y_u^{(1 \rightarrow 2)}} ), \quad &q_u^{(2)} \ge 2 ~\text{and}~ \frac{q_u^{(1)}}{q_u^{(2)}} \in \mathbb{N} \\
 f_y ~\text{as in (\ref{eq:ypdf})} , \quad & \text{otherwise}
\end{cases}
\end{equation}
Note that the distribution of $y_{u}^{(2)}$ for the first two cases depends on the variance of the auxiliary noise $y_{u}^{(1 \rightarrow 2)}$.
\end{proposition}

Note that Proposition~\ref{pro:secondq} is rather different from Property~\ref{pro:round1q} of the first-cycle quantization noise.

\vspace{0.3cm}
\noindent \textbf{Rounding noise:} For rounding noise, we look at two cases. This first one corresponds to the divisible quantization condition, and the second one covers the remaining possibilities. The following proposition gives the distribution for the second-cycle rounding noise.

\begin{proposition} \label{pro:secondr} For all spatial index $m$, the second-cycle rounding noise has the following distributions.
\begin{equation}\label{eq:secondr}
    x_m^{(2 \rightarrow 3)}  \sim
        \begin{cases}
        f_{x_m^{(1 \rightarrow 2)}}, &q_u^{(2)} \ge 2 ~\text{and}~ \frac{q_u^{(1)}}{q_u^{(2)}} \in \mathbb{N}, \forall u  \\
        \mathcal{Q^N}( \sigma^2_{x_m^{(2)}}   ,1 ), &\text{otherwise} \\
        \end{cases}
\end{equation}
where $f_{x_m^{(1 \rightarrow 2)}}$ denotes the probability density function for $x_m^{(1 \rightarrow 2)}$.

It is clear that the rounding noise follows the quantized-Gaussian distribution.
Note that when the divisible quantization condition is satisfied, the rounding noise distribution is inherited directly from the previous cycle without change.
Otherwise, the distribution parameter may change.

\end{proposition}

The rounding noise distribution of the first cycle is given by Property~\ref{pro:round1r}. Comparing Property~\ref{pro:round1r} of the first-cycle rounding noise with Proposition~\ref{pro:secondr} , the former is a specialization of the latter.

\subsection{JPEG noises of Higher Compression Cycle} \label{ssec:thirdcycle}

The results for the second compression cycle can be easily generalized for both quantization noise as in Proposition~\ref{pro:kcycleq} and rounding noise as in Proposition~\ref{pro:kcycler}. The proof of these propositions can be found in Appendix.

\begin{proposition} \label{pro:kcycleq} The ($k$+$1$)-cycle quantization noise (where $k \ge 2$) follows the following distributions, for all DCT coefficient index $u$:

\begin{equation}\label{eq:kcycleq}
\begin{split}
y_u^{(k+1)} \sim
\begin{cases}
 \mathcal{Q^N}( \sigma^2_{y_u^{(k \rightarrow k+1)}}   ,1), \quad &q_u^{(k+1)} = 1  \\
 \mathcal{N}( 0, \sigma^2_{y_u^{(k \rightarrow k+1)}} ), \quad &q_u^{(k+1)} \ge 2 ~\text{and}~ \frac{q_u^{(k)}}{q_u^{(k+1)}} \in \mathbb{N} \\
 f_y ~\text{as in (\ref{eq:ypdf})} , \quad & \text{otherwise}
\end{cases}
\end{split}
\end{equation}
Note that the distribution of $y_{u}^{(k+1)}$ for the first two cases depends on the variance of the auxiliary noise $y_{u}^{(k \rightarrow k+1)}$.
\end{proposition}

\begin{proposition} \label{pro:kcycler} The ($k$+$1$)-cycle rounding noise (where $k \ge 2$) follows the quantized Gaussian distributions. For all spatial index $m$, that is,
\begin{equation}\label{eq:kcycler}
    x_m^{(k+1 \rightarrow k+2)}  \sim
        \begin{cases}
        f_{x_m^{(k \rightarrow k+1)}}, &q_u^{(k+1)} \ge 2, \frac{q_u^{(k)}}{q_u^{(K+1)}} \in \mathbb{N}, \forall $u$  \\
        \mathcal{Q^N}( \sigma^2_{x_m^{(k+1)}}   ,1 ), &\text{otherwise} \\
        \end{cases}
\end{equation}
Note that when the first condition is satisfied, the rounding noise distribution is inherited directly from the previous cycle without change. Otherwise, the distribution parameter may change.
\end{proposition}

\section{Applications}

The significance of the results for higher compression cycle obtained in the previous section goes beyond mere theoretical interest. It leads to new algorithms with superior results. We show results for two applications of JPEG noises. The first one is JPEG quantization step estimation that utilizes the quantization noise and the second one is JPEG identical re-compression detection that makes use of the rounding noise.

We show experimental results that is enough to illustrate the case of application for our theoretical findings. For detailed description of the algorithms and more comprehensive results, we refer readers to~\cite{binTechReport1} and \cite{binTechReport2}.

\subsection{JPEG Quantization Step Estimation}

JPEG quantization step estimation is important for image forensics, JPEG image enhancement, and image quality assessment.

\vspace{0.3cm}
\noindent \textbf{Problem Statement:} JPEG quantization step estimation refers to a problem of estimating quantization steps of the first-cycle compression given an decompressed JPEG image, where all quantization metainfo is stripped away. Our goal is to estimate $q^{(1)}$ from the decompressed JPEG image.

\vspace{0.3cm}
\noindent \textbf{Methodology:} Although the first-cycle quantization noise $y^{(1)}$ can no longer be observed, we can compute the second-cycle quantization noise $y^{(2)}$ for various quantization step sizes $q^{(2)}$ from an decompressed JPEG image. We can show that $y^{(2)}$, which is a function of $q^{(2)}$, is closely related to $q^{(1)}$. We can see that when $q^{(1)} = q^{(2)}$, the divisible quantization condition is satisfied as long as $q^{(2)} \ge 2$. In this case, Proposition \ref{pro:secondq} tells us that $y^{(2)}$ will be a zero-mean Gaussian. As a result, we can study the relevant property of $y^{(2)}$ through its sample variance, which is also its sufficient statistics.

\begin{equation}\label{eq:Nstatistics}
  S^{var}(q) = \frac{1}{N}\sum_N  \left(  y_{}^{(2)}(q) \right)^2 , \quad {q} \in \mathbb{N},
\end{equation}
where $N$ is the number of noise samples.

As a main result, Proposition~\ref{pro:minimum} shows us that when $q^{(2)}$ is equal to the true quantization step size of the first cycle, a local minimum appears on the curve given by $S^{var}(q)$ the sample variance of $y^{(2)}$ as a function of $q^{(2)}$.

\begin{proposition} \label{pro:minimum}
    Let $\dot{q} \in \mathbb{N}$ be an integer divisor of the true quantization step $q^*$ ($\dot{q}$ is hence no larger than $q^*$),
    and $\ddot{q} \in \mathbb{N}$ be an integer other than the above.
    The following relation holds:
    \begin{equation}\label{eq:relationP-A}
            Var \big[ y_{}^{(2)}({  \dot{q}  }) \big]   \le
            Var \big[ y_{}^{(2)}({  q^*  }) \big] <
            Var \big[ y_{}^{(2)}({  \ddot{q}  }) \big],
    \end{equation}
    where $q^* \in \mathbb{N} $ and $Var[\cdot]$ is the variance of a random variance.
\end{proposition}

The local minimum property enables us to identify the true quantization step size by searching for local minima on $S^{var}(q)$. This step reduces the search space to a finite set of discrete locations that correspond to the local minima on $S^{var}(q)$. To identify the true quantization step, a further step of eliminating invalid local minima is needed. We refer readers to~\cite{binTechReport1} for the details of the local minima elimination step.

As a result, our algorithm which begins with a local minima searching step that followed by a local minima elimination step can be succinctly described below:

\begin{algorithm} \label{algo:A}
The true quantization step size $q^*$ of the first-cycle JPEG compression can be estimated as follows when the parameters $T_c$ and $T_{\xi}$ are given:
 \begin{equation}\label{eq:algo-A}
	q^* =
	\begin{cases}
	\hat{q}, \quad &\hat{q} \ge 2  \\
 	2, \quad &\hat{q} < 2 \text{ and } S^{var}(2) < T_c \\
 	1, \quad &\hat{q} < 2  \text{ and } S^{var}(2) \ge T_c,
\end{cases}
\end{equation}
    where $\hat{q} = \arg \max_{q \in \mathbb{N}} \{ L_{min} [ S^{var}(q) ] ~|~ S^{var}(q) <T_{\xi} \}$, and $L_{min}[\cdot]$ denotes the local minima search function.
\end{algorithm}

We refer readers to~\cite{binTechReport1} for the details on selecting the optimal value for parameters $T_c$ and $T_{\xi}$.

\vspace{0.3cm}
\noindent \textbf{Experiment:} In our experiment, we evaluate how well our method perform for estimating quantization step from JPEG decompressed images. For this experiment, we simulated JPEG compression using quantization tables with a constant number, chosen from a set $\{1$, $2$, $3$, $4$, $5$, $6$, $7$, $10$, $13$, $16$, $19$, $22$, $25$,    $28$,    $31$,    $34$,    $37$,    $40$,    $43$,    $46$, $49$,    $52$,    $55$,    $58$, $61\}$. We define the estimation accuracy in this experiment as the percentage of correct estimation over all trials. A number of 3000 images are used for generating simulation images. The 3000 images are collected from three independent sources \cite{Schaefer2003,USDA2012,Bas2011} with equal number of images from them.

We compare our method, denoted by \textbf{L2C}, to four other methods with state-of-the-art performance.

\begin{enumerate}
  \item \textbf{EDS method}: A JPEG coefficient method by Lin \emph{et al.} \cite{Lin2011} that is based on the energy density spectrum of the histogram of DCT coefficients.
  \item \textbf{PEAK method}: A JPEG coefficient method by Luo \emph{et al}. \cite{Luo2010} that regards the bin index of the first peak in the histogram of non-zero rounded DCT coefficients as the estimated quantization step.
  \item \textbf{MLE method}: A JPEG quantization noise method by Neelamani  \emph{et al.} \cite{Neelamani2006} that is based on a maximum likelihood estimator. We use the implementation obtained from the authors~\cite{mlecode}.
  \item \textbf{L1A method}: A JPEG quantization noise method by Fridrich \emph{et al.} \cite{Fridrich2001} that is based on an $L1$-norm noise statistics.
\end{enumerate}

\begin{figure*}[!t]
\centering
\subfigure[Image size: $256 \times 256$]{\includegraphics[width=3.2in]{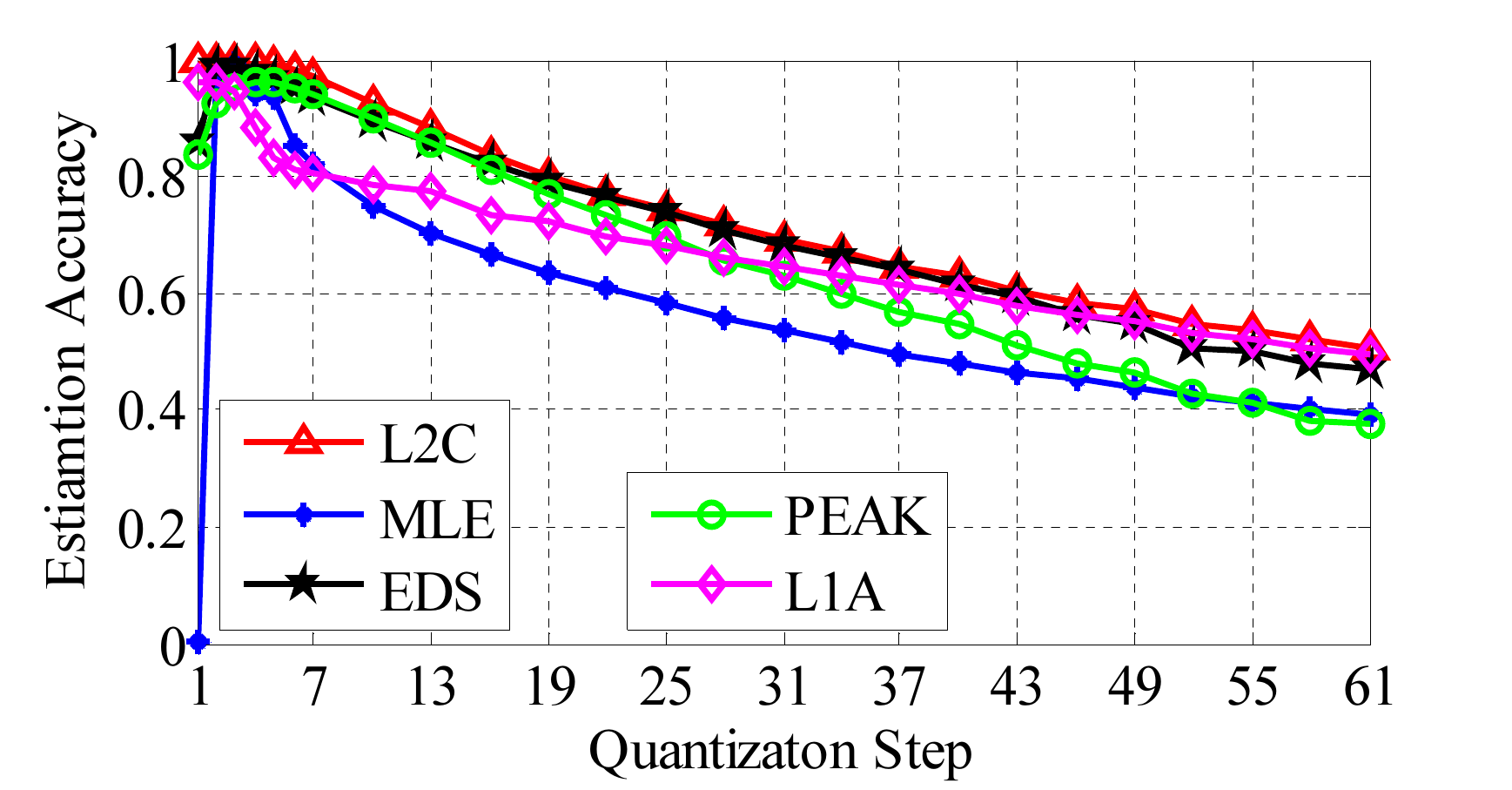}}
\subfigure[Image size: $128 \times 128$]{\includegraphics[width=3.2in]{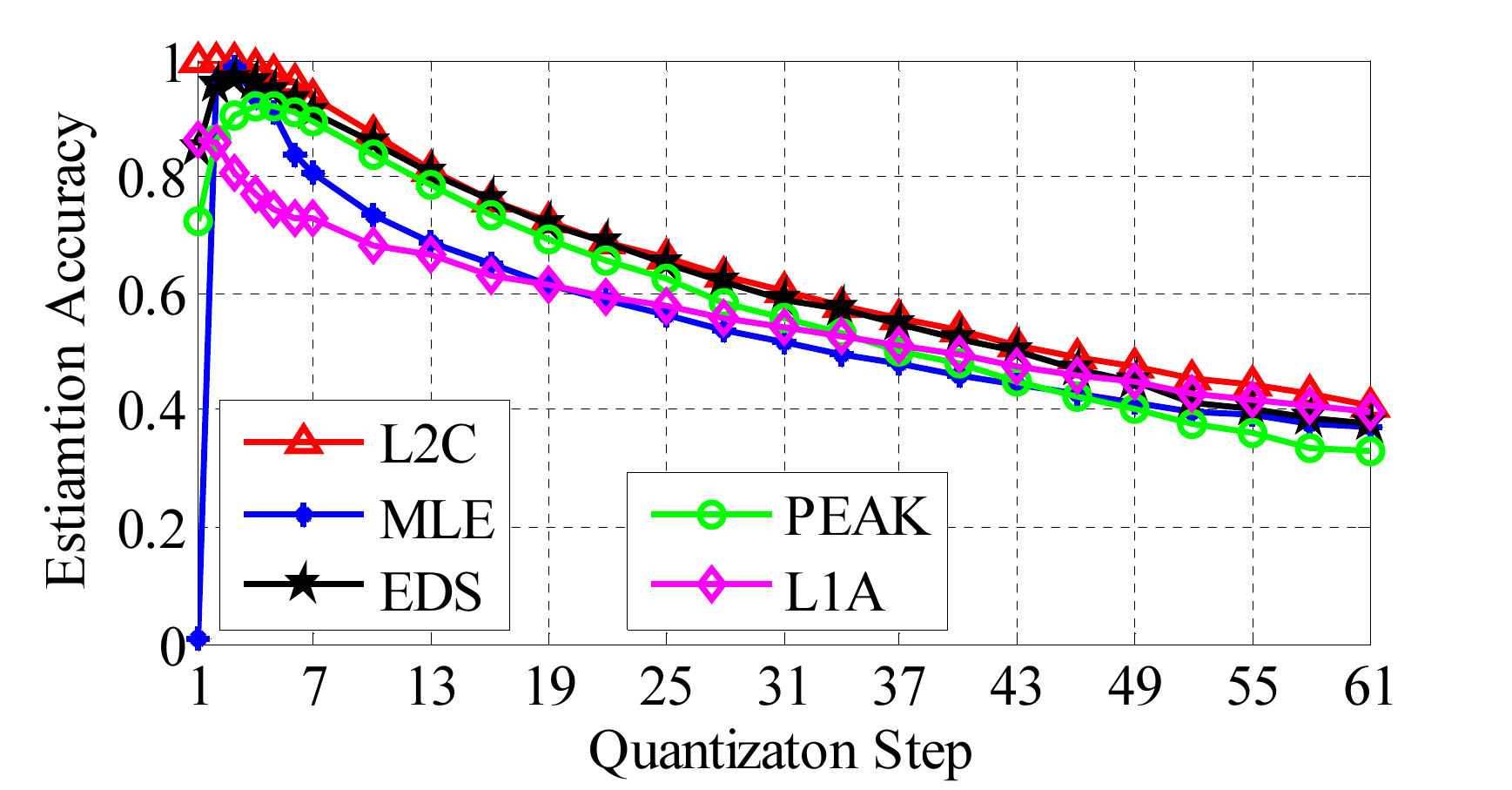}}
\subfigure[Image size: $64 \times 64$]{\includegraphics[width=3.2in]{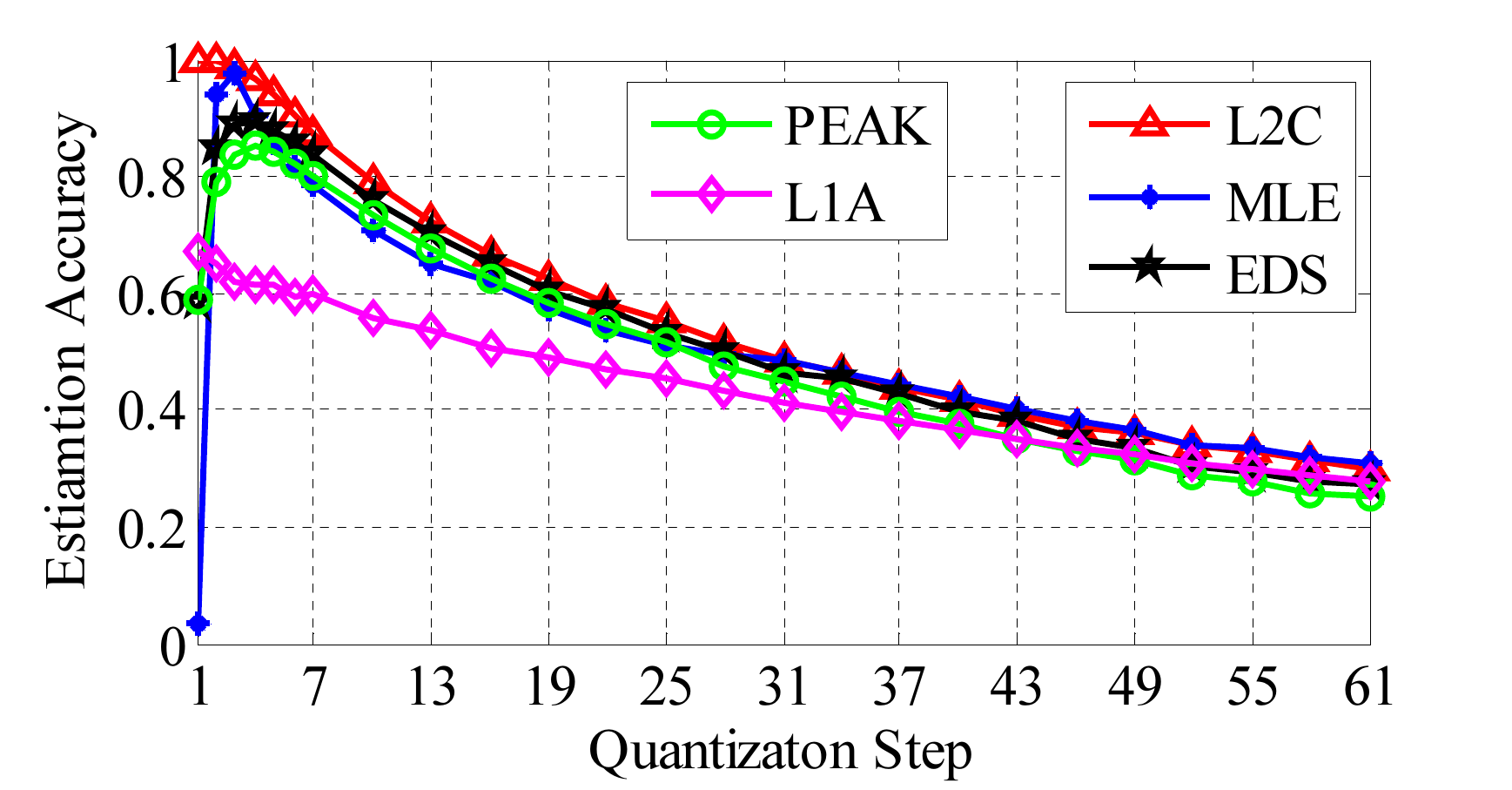}}
\subfigure[Image size: $32 \times 32$]{\includegraphics[width=3.2in]{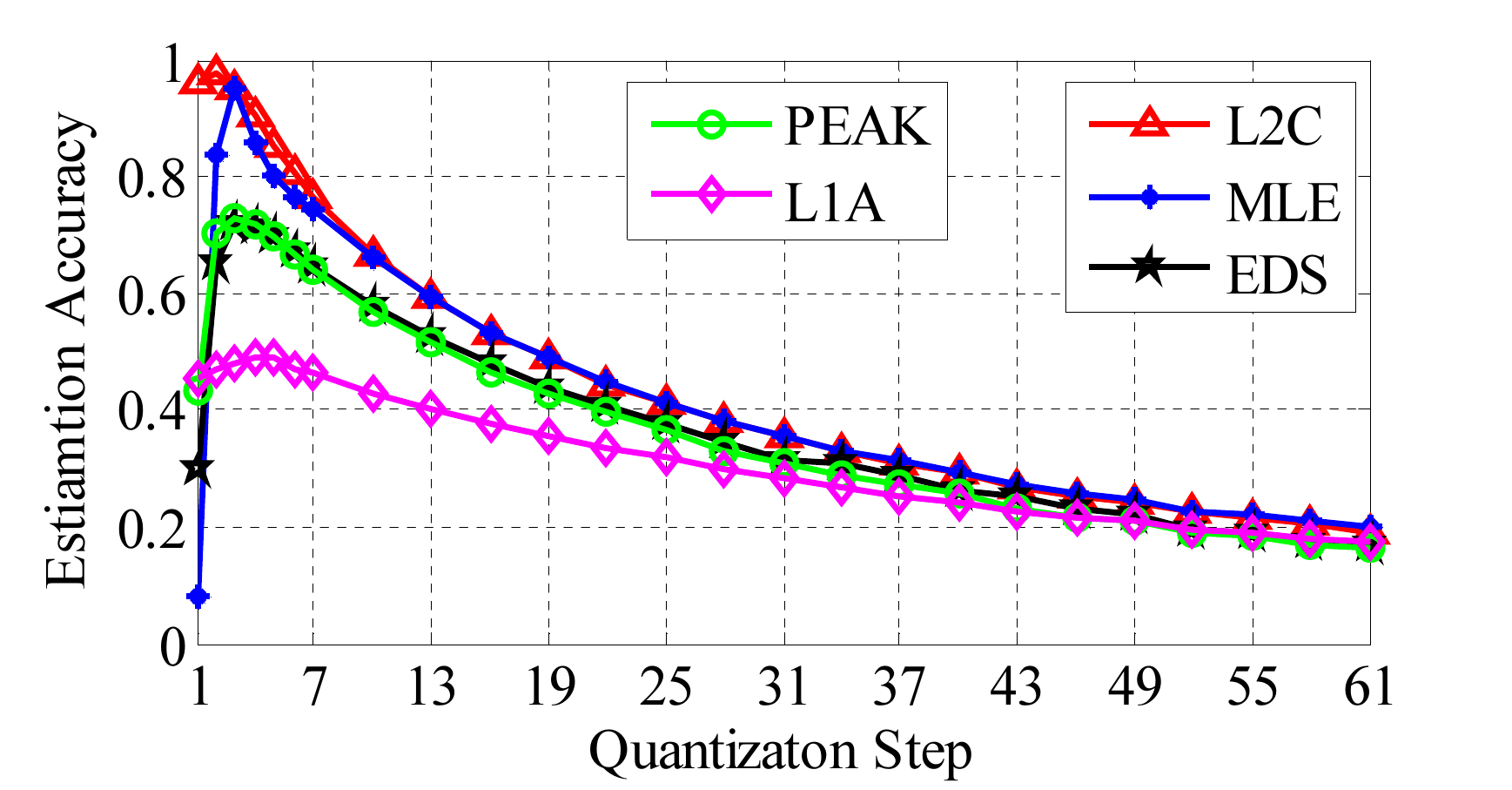}}
\caption{Quantization step estimation performance when quantization table containing the same step size. Due to possible low printing resolution, readers are encouraged to enlarge the figure on a computer screen for better viewing.}\label{fig_idest}
\end{figure*}

The results are shown in Fig. \ref{fig_idest}. We can see that our method outperforms the competing methods, especially for small quantization steps by a large margin. The results are demonstrated in Fig. \ref{fig_idest}. We can observe that the proposed method performs the best. Unlike other methods, the performance of our method increases monotonically as image size increases and quantization step size decreases. This is an indication that the performance of our method is consistent with the amount of quantization information available in an image.

Generally, the JPEG coefficient methods (EDS and PEAK) perform better than the JPEG quantization noise methods (MLE and L1A) when the image size is large, but they perform worse than MLE when the image size is small.
The reason may due to the fact that the coefficient-based methods require more coefficient samples to construct a reliable histogram, hence these methods are not suitable for small-size images.
MLE, EDS, and PEAK do not perform well for the case when the true quantization step size is $1$.
This is because these methods either cannot differentiate uncompressed  images and the high-quality compressed images, or may be confused about whether the true quantization step size is $1$ or $2$.
The poor performance obtained by L1A is due to the loose threshold that is not effective in isolating non-quantization-induced local minima, which appear at the multiples of the true quantization step.

\subsection{JPEG Identical Re-compression Identification}

In general, a doubly-compressed JPEG image can be easily identified due to the abnormal distribution of DCT coefficients, which is caused by quantization with different steps in two successive compressive cycles. However, a re-compressed JPEG image with identical quantization parameters in two successive cycles is often deemed indistinguishable from a singly-compressed image~\cite{mahdian2009detecting}. To our knowledge, Huang \textit{et. al.}'s method \cite{Huang2010} is the only work addressing identical re-compression problem. However, this method lacks theoretical grounding and is sensitive to parameter setting.

\vspace{0.3cm}
\noindent \textbf{Problem Statement:} JPEG identical re-compression identification refers to a problem of identifying whether an image in JPEG format has been compressed once or compressed twice with the same quantization table. In this case, we assume that the quantization table can be extracted from the JPEG metadata.

To detect identical re-compression, we compute the rounding noise variance. In the context to Fig. \ref{fig:flowchart2}, the rounding noise computed would be
$x_{m}^{(1 \rightarrow 2)}$ if the given image is only compressed once. Otherwise, it would be $x_{m}^{(2 \rightarrow 3)}$. Due to identical compression, we assume the quantization table of the first cycle is the same as the second cycle, \emph{i.e.}, $q_u^{(1)} = q_u^{(2)}$ for all DCT coefficient index $u$. Under this context, our goal is to distinguish $x_{m}^{(1 \rightarrow 2)}$ from $x_{m}^{(2 \rightarrow 3)}$.

\vspace{0.3cm}
\noindent \textbf{Methodology:} Based on Property \ref{pro:round1r} and Proposition \ref{pro:secondr}, we know that the rounding noise follows a quantized Gaussian distribution. When the divisible quantization condition is satisfied, $x_{m}^{(1 \rightarrow 2)}$ and $x_{m}^{(2 \rightarrow 3)}$ will be identically distributed and hence statistically similar. To differentiate the two, we need to consider situations where the divisible quantization condition fails to hold, which happens when $q^{(2)} = 1$. Under this circumstances, our main result, Proposition~\ref{pro:id2}, says more about $x_{m}^{(1 \rightarrow 2)}$ and $x_{m}^{(2 \rightarrow 3)}$ than that they are different:

\begin{proposition} \label{pro:id2} When the divisible quantization condition does not hold in the context of JPEG identical re-compression, we have the following:
\begin{equation}\label{eq:id2}
	 \sigma^2_{x_m^{(2 \rightarrow 3)}}     \leq
     \sigma^2_{x_m^{(1 \rightarrow 2)}},  ~~\forall m.
\end{equation}
\end{proposition}

With Proposition~\ref{pro:id2}, we propose a very simple algorithm for detecting JPEG identical re-compression:

\begin{algorithm} \label{algo:B}
JPEG identical re-compression can be detection as follows when the parameter $T$ is given:
 \begin{equation}\label{eq:algo-A}
	\text{Image} =
	\begin{cases}
	\text{Singly compressed}, \quad & \sigma^{2}_{x_{all}} > T  \\
 	\text{Identically double-compressed}, \quad & \sigma^{2}_{x_{all}} \le T,
\end{cases}
\end{equation}
    where $\sigma^{2}_{x_{all}}$ is variance of rounding noise for all spatial locations.
\end{algorithm}

We refer readers to~\cite{binTechReport2} for the details on how to select a good value for parameter $T$.

\vspace{0.3cm}
\noindent \textbf{Experiment:} In this experiment, we evaluate the capability of our simple method on detecting identically recompressed JPEG images. We compare our method with Huang \textit{et. al.}'s method on 3000 simulation images generated from uncompressed images of independent sources \cite{Schaefer2003,USDA2012,Bas2011}. We reserve 1000 images for setting parameters in both our method and Huang's, and use the rest images for testing.
The images are compressed by using quality factor (QF) compatible to IJG (Independent JPEG Group). The results are shown in Table \ref{tab:id_double_jpegC} for different image size ranging from $128 \times 128$ to $16 \times 16$.

The classification results are reported in Table \ref{tab:id_double_jpegA} to \ref{tab:id_double_jpegC},
where the accuracy is the average of true positive rate (correctly classified as identical re-compressed image) and true negative rate (correctly classified as singly compressed image).

It can be observed that for QF that is is no less than $95$, the performance of our method is comparable to Huang's method and outperforms Huang's when the image size is small.
  However, both methods suffer sudden performance drop around QF=93.
  When QF is no larger than $92$ as all entries in the quantization table are greater than 1, our method fails as the QF is outside of the detectable domain for our method. In this domain, Huang's method performs slightly better than random guessing. This gives a good reason for fusing our method and Huang's method to achieve a more powerful detector for identical re-compression.

\renewcommand{\arraystretch}{0.8}
\begin{table*}[!htb]
\caption{Accuracy of Classifying Singly Compressed Images and Identical re-compressed Images on Set A.}
\label{tab:id_double_jpegA}
  \centering
  \begin{tabular}{c c c    c c c   c c c}
    \toprule
     {IJG } &
     \multicolumn{2}{c}{$128 \times 128$} & \multicolumn{2}{c}{$64 \times 64$} & \multicolumn{2}{c}{$32 \times 32$} & \multicolumn{2}{c}{$16 \times 16$} \\
     \cmidrule(r){2-3}  \cmidrule(r){4-5}  \cmidrule(r){6-7}  \cmidrule(r){8-9}
     QF & Huang's & Ours &  Huang's & Ours &  Huang's & Ours & Huang's & Ours \\
    \midrule
100  &	\textbf{99.98} 	&	99.73 	&	99.43 	&	\textbf{99.73} 	&	92.28 	&	\textbf{99.65} 	&	78.48 	&	\textbf{96.98} 	\\
99	&	\textbf{99.98} 	&	99.85 	&	\textbf{99.90} 	&	99.85 	&	95.38 	&	\textbf{99.70} 	&	83.00 	&	\textbf{98.30} 	\\
98	&	\textbf{99.9}3 	&	99.68 	&	99.08 	&	\textbf{99.73} 	&	92.40 	&	\textbf{99.68} 	&	77.60 	&	\textbf{94.68} 	\\
97	&	\textbf{99.75} 	&	99.60 	&	97.45 	&	\textbf{99.70} 	&	90.50 	&	\textbf{99.48} 	&	75.68 	&	\textbf{92.33} 	\\
96	&	99.43 	&	\textbf{99.55} 	&	95.75 	&	\textbf{99.58} 	&	88.10 	&	\textbf{98.40} 	&	73.58 	&	\textbf{86.75} 	\\
95	&	99.30 	&	\textbf{99.40} 	&	96.65 	&	\textbf{99.23} 	&	87.10 	&	\textbf{92.83} 	&	73.13 	&	\textbf{77.40} 	\\
94	&	\textbf{99.43} 	&	98.95 	&	\textbf{95.23} 	&	94.13 	&	\textbf{84.90} 	&	80.48 	&	\textbf{71.65} 	&	65.65 	\\
93	&	\textbf{97.63} 	&	81.85 	&	\textbf{89.98} 	&	68.48 	&	\textbf{78.18} 	&	60.75 	&	\textbf{64.33} 	&	54.88 	\\
92	&	\textbf{83.10} 	&	56.78 	&	\textbf{66.95} 	&	53.28 	&	\textbf{55.90} 	&	51.60 	&	\textbf{52.35} 	&	51.15 	\\
91	&	\textbf{77.88} 	&	55.45 	&	\textbf{63.35} 	&	52.28 	&	\textbf{54.58} 	&	51.48 	&	\textbf{51.75} 	&	50.50 	\\
90	&	\textbf{74.93} 	&	54.95 	&	\textbf{61.65} 	&	52.60 	&	\textbf{54.25} 	&	51.08 	&	\textbf{51.55} 	&	50.48 	\\
85	&	\textbf{61.28} 	&	52.38 	&	\textbf{56.10} 	&	50.93 	&	\textbf{52.40} 	&	50.35 	&	\textbf{50.93} 	&	50.05 	\\
75	&	\textbf{58.23} 	&	50.00 	&	\textbf{53.93} 	&	50.00 	&	\textbf{51.80} 	&	49.98 	&	\textbf{50.88} 	&	49.90 	\\
    \bottomrule
  \end{tabular}
\end{table*}
\renewcommand{\arraystretch}{0.8}
\begin{table*}[!htb]
\caption{Accuracy of Classifying Singly Compressed Images and Identical re-compressed Images on Set B.}
\label{tab:id_double_jpegB}
  \centering
  \begin{tabular}{c c c    c c c   c c c}
    \toprule
     {IJG } &
     \multicolumn{2}{c}{$128 \times 128$} & \multicolumn{2}{c}{$64 \times 64$} & \multicolumn{2}{c}{$32 \times 32$} & \multicolumn{2}{c}{$16 \times 16$} \\
     \cmidrule(r){2-3}  \cmidrule(r){4-5}  \cmidrule(r){6-7}  \cmidrule(r){8-9}
     QF & Huang's & Ours &  Huang's & Ours &  Huang's & Ours & Huang's & Ours \\
    \midrule
100     &	\textbf{99.70} 	&	99.60 	&	99.08 	&	\textbf{99.70} 	&	93.48 	&	\textbf{99.73} 	&	78.03 	&	\textbf{96.88} 	\\
99	&	99.45 	&	\textbf{99.75} 	&	99.20 	&	\textbf{99.75} 	&	95.70 	&	\textbf{99.73} 	&	82.85 	&	\textbf{98.25} 	\\
98	&	\textbf{99.68} 	&	99.53 	&	97.08 	&	\textbf{99.63} 	&	90.95 	&	\textbf{99.50} 	&	75.60 	&	\textbf{94.70} 	\\
97	&	\textbf{99.58} 	&	99.48 	&	98.10 	&	\textbf{99.60} 	&	89.35 	&	\textbf{99.28} 	&	75.35 	&	\textbf{92.25} 	\\
96	&	99.05 	&	\textbf{99.40} 	&	96.05 	&	\textbf{99.40} 	&	86.15 	&	\textbf{98.13} 	&	74.58 	&	\textbf{86.23} 	\\
95	&	97.90 	&	\textbf{99.15} 	&	94.88 	&	\textbf{99.10} 	&	85.53 	&	\textbf{92.85} 	&	73.45 	&	\textbf{77.58} 	\\
94	&	98.20 	&	\textbf{98.50} 	&	\textbf{94.73} 	&	94.55 	&	\textbf{84.28} 	&	79.65 	&	\textbf{72.45} 	&	65.70 	\\
93	&	\textbf{96.43} 	&	81.05 	&	\textbf{89.85} 	&	69.30 	&	\textbf{79.40} 	&	60.40 	&	\textbf{64.60} 	&	55.48 	\\
92	&	\textbf{85.50} 	&	57.58 	&	\textbf{68.53} 	&	54.23 	&	\textbf{57.40} 	&	51.98 	&	\textbf{52.40} 	&	51.30 	\\
91	&	\textbf{80.45} 	&	56.45 	&	\textbf{66.03} 	&	53.53 	&	\textbf{56.30} 	&	51.95 	&	\textbf{52.55} 	&	51.08 	\\
90	&	\textbf{77.90} 	&	55.93 	&	\textbf{64.48} 	&	53.43 	&	\textbf{55.45} 	&	51.43 	&	\textbf{52.15} 	&	50.88 	\\
85	&	\textbf{66.85} 	&	55.00 	&	\textbf{59.25} 	&	52.05 	&	\textbf{54.03} 	&	51.33 	&	\textbf{51.75} 	&	50.48 	\\
75	&	\textbf{57.13} 	&	50.08 	&	\textbf{53.23} 	&	49.88 	&	\textbf{51.38} 	&	49.88 	&	\textbf{50.65} 	&	50.00 	\\
    \bottomrule
  \end{tabular}
\end{table*}
\renewcommand{\arraystretch}{0.8}
\begin{table*}[!htb]
\caption{Accuracy of Classifying Singly Compressed Images and Identical re-compressed Images on Set C.}
\label{tab:id_double_jpegC}
  \centering
  \begin{tabular}{c c c  c c c  c c c}
    \toprule
     {IJG } &
     \multicolumn{2}{c}{$128 \times 128$} & \multicolumn{2}{c}{$64 \times 64$} & \multicolumn{2}{c}{$32 \times 32$} & \multicolumn{2}{c}{$16 \times 16$} \\
     \cmidrule(r){2-3}  \cmidrule(r){4-5}  \cmidrule(r){6-7}  \cmidrule(r){8-9}
     QF & Huang's & Ours &  Huang's & Ours &  Huang's & Ours & Huang's & Ours \\
    \midrule
100	    &	\textbf{99.88} 	&	99.65 	&	99.43 	&	\textbf{99.68} 	&	91.90 	&	\textbf{99.73} 	&	77.48 	&	\textbf{96.88} 	\\
99	&	99.73 	&	\textbf{99.75} 	&	99.68 	&	\textbf{99.75}	&	96.43 	&	\textbf{99.70} 	&	83.63 	&	\textbf{98.58} 	\\
98	&	99.50 	&	\textbf{99.65} 	&	98.55 	&	\textbf{99.58} 	&	91.00 	&	\textbf{99.60} 	&	76.98 	&	\textbf{94.48} 	\\

97	&\textbf{99.65} 	&	99.63 	&	98.30 	&	\textbf{99.65} 	&	88.33 	&\textbf{99.50} 	&	75.83 	&	\textbf{92.15} 	\\

96	&	98.90 	&	\textbf{99.60} 	&	96.63 	&	\textbf{99.55} 	&	85.85 	&\textbf{98.33} 	&	73.98 	&	\textbf{86.43} 	\\

95	&	98.63 	&	\textbf{99.45} 	&	95.73 	&	\textbf{99.28} 	&	85.83 	&	\textbf{92.58} 	&	72.73 	&	\textbf{77.13} 	\\
94	&	98.28 	&	\textbf{99.05} 	&	\textbf{95.20}	&	93.98 	&	\textbf{84.25} 	&	80.28 	&	\textbf{71.90} 	&	65.93 	\\
93	&	\textbf{96.65} 	&	81.75 	&	\textbf{89.28} 	&	69.03 	&	\textbf{80.00} 	&	60.75 	&	\textbf{64.15} 	&	55.23 	\\
92	&	\textbf{85.78} 	&	56.75 	&\textbf{	68.55 }	&	52.50 	&	\textbf{57.05} 	&	52.03 	&	\textbf{52.33} 	&	51.28 	\\
91	&	\textbf{79.95} 	&	55.05 	&\textbf{	63.73 }	&	52.55 	&	\textbf{55.13} 	&	51.58 	&	\textbf{52.00} 	&	50.93 	\\
90	&	\textbf{77.18} 	&	54.48 	&	\textbf{62.75} 	&	52.93 	&	\textbf{55.05} 	&	51.38 	&	\textbf{51.78} 	&	50.88 	\\
85	&	\textbf{62.60} 	&	54.13 	&	\textbf{57.00} 	&	51.98 	&	\textbf{53.20} 	&	51.30 	&	\textbf{51.35} 	&	50.58 	\\
75	&	\textbf{54.43} 	&	50.10 	&	\textbf{52.08} 	&	50.03 	&	\textbf{50.95} 	&	50.05 	&	\textbf{50.43} 	&	50.00 	\\
    \bottomrule
  \end{tabular}
\end{table*}

\section{Conclusion}

In this paper, we present a statistical analysis on JPEG noises beyond the first compression cycle.
We show that the noise distributions in higher compression cycles are rather different from those in the first compression cycle.
Specifically, for the first compression cycle, the quantization noise in the DC component follows a uniform distribution, and that in the AC component follows a quantized Laplacian distribution.
The rounding noise follows a quantized-Gaussian distribution.
For higher compression cycles, the quantization noise may follow a quantized-Gaussian distribution, or a Gaussian distribution, or a distribution characterized by the distribution of DCT coefficients.
The rounding noise may follow the same distribution as that in the previous compression cycle, or a quantized-Gaussian distribution with a different variance to previous one.
It is the quantization parameters that determines the type of the distribution.

Since the distribution models are parameterized by the quantization steps involved in each compression cycle,
they can help to uncover JPEG compression history.
Two applications of the noise distribution model,  JPEG quantization step estimation and JPEG re-compression identification, are presented in this paper.
However, we believe that the applications are more than that.
For example, the analytical results here may also be extended to other block-DCT based coding schemes, such MPEG.
We regard them as our future work.

\appendices

\section{Justification of Proposition \ref{pro:secondq} }

As the statements are independent of the DCT coefficient index $u$, we drop the index for clarity.

By \eqref{eq:Nq1} and \eqref{eq:Ny}, we know that for $k \ge 1$:
\begin{equation}\label{eq:y_relationA1}
\begin{split}
{y}_{}^{(k+1)}
&=  {{Y}_{ }}^{(k+1)} - \left[  \frac{ {{Y}_{ }}^{(k+1)}} {q_{ }^{(k+1)}  } \right]{q_{ }^{(k+1)}} \\
&=   {\tilde{Y}_{ }}^{(k)} -  {y}_{ }^{(k \rightarrow k+1)}  - \left [   \frac{ {\tilde{Y}}_{ }^{(k)} -  {y}_{ }^{(k \rightarrow k+1)} }{q_{ }^{(k+1)}} \right]{q_{ }^{(k+1)}} \\
\end{split}
\end{equation}
Notice that ${\tilde{Y}_{ }}^{(k)}$ is an integer multiple of ${q}_{ }^{(k)}$.
When  ${{q}_{ }^{(k)}}$ is also an integer multiple of ${{q}_{ }^{(k+1)}}$,
 ${\tilde{Y}_{ }}^{(k)}$ is still an integer multiple of ${q}_{ }^{(k+1)}$, i.e.,
\begin{equation}\label{eq:y_relationB1}
{ {\tilde{Y}_{ }}^{(k)}}
 =  \left[  \frac { {\tilde{Y}_{ }}^{(k)}}{q_{ }^{(k+1)}}  \right] {q_{ }^{(k+1)}},  \quad \frac{{q}_{ }^{(k)}} {{q}_{ }^{(k+1)}}  \in \mathbb{N}
\end{equation}
Using the expression \eqref{eq:y_relationB1} in \eqref{eq:y_relationA1}, we can derive
\begin{equation}\label{eq:y_relationC1}
\begin{split}
{y}_{ }^{(k+1)}
&=  \left( \left[  \frac { {\tilde{Y}_{ }}^{(k)}}{q_{ }^{(k+1)}}  \right]  -
 \left [   \frac{ {\tilde{Y}}_{ }^{(k)} -  {y}_{ }^{(k \rightarrow k+1)} }{q_{ }^{(k+1)}} \right] \right)
 {q_{ }^{(k+1)}} \\
 &-  {y}_{ }^{(k \rightarrow k+1)}  \\
&= -\left (  {y}_{ }^{(k \rightarrow k+1)} -  \left [ \frac{ {y}_{ }^{(k \rightarrow k+1)}}{q_{ }^{(k+1)}} \right] {q_{ }^{(k+1)}}  \right)
\end{split}
\end{equation}
The above equation implies that when $\frac{{q}_{ }^{(k)}} {{q}_{ }^{(k+1)}}  \in \mathbb{N}$, we can regard ${y}_{ }^{(k+1)}$ as negatively quantizing ${y}_{ }^{(k \rightarrow k+1)}$ with quantization step ${q}_{ }^{(k+1)}$.

For $k=1$, we have the following three cases.
\begin{enumerate}
  \item[(a)]
When $ {q}_{}^{(2)} = {1}$, 
\eqref{eq:y_relationC1} is reduced to
\begin{equation}\label{eq:y_relationD1}
{y}_{ }^{(2)}
= - \left(  {y}_{ }^{(1 \rightarrow 2)} -   [ { {y}_{ }^{(1 \rightarrow 2)}} ]    \right)
\end{equation}
Since the auxiliary noise $ {y}_{ }^{(1 \rightarrow 2)}$ $\sim$ $\mathcal{N}(0, \sigma^2_{{y}_{ }^{(1 \rightarrow 2)}})$, 
hence we arrive ${y}_{ }^{(2)}$ $\sim$ $\mathcal{Q^N}(\sigma^2_{{y}_{ }^{(1 \rightarrow 2)}},1)$.

  \item[(b)]
When ${q}_{}^{(2)} \ge 2 $ and $\frac{q_{ }^{(1)}}{q_{ }^{(2)}} \in \mathbb{N} $,
based on the variance bounds shown in \eqref{eq:round1r_bound} and \eqref{eq:round1aux_bound2},
we have $\sigma_{  y_{ }^{(1 \rightarrow 2)}  } \le \sqrt{1/12} = 0.289$.
Note that for a random variable $Z \sim \mathcal{N}(0, \sigma_{Z}^{2})$, the probability $\Pr \{ | Z| < 3 \sigma_{Z}   \} \approx 1$.
Therefore, we have
 \begin{equation}\label{eq:3sigma}
    \begin{split}
         \Pr \left\{  \left[ \frac{y_{ }^{(1 \rightarrow 2)}}{q_{ }^{(2)}} \right] = 0  \right\}
         &= \Pr  \left\{    {y_{ }^{(1 \rightarrow 2)}} < \frac{1}{2} {q_{ }^{(2)}}  \right \}  \\
         & > \Pr  \left\{    \left | \frac{y_{ }^{(1 \rightarrow 2)}}{q_{ }^{(2)}} \right | < \frac{3}{\sqrt{12} }   \right \}
         \approx 1
     \end{split}
 \end{equation}
Applying the relation \eqref{eq:3sigma} in \eqref{eq:y_relationC1}, we have
 \begin{equation}\label{eq:jpeg_APR_relationA1}
 \Pr \left \{  y_{ }^{(2)}   = - {y_{ }^{(1 \rightarrow 2)}} \right \} \approx 1
 \end{equation}
Since the auxiliary noise $ {y}_{ }^{(1 \rightarrow 2)}$ $\sim$ $\mathcal{N}(0, \sigma^2_{{y}_{ }^{(1 \rightarrow 2)}})$,
we arrive $ y_{ }^{(2)} \sim  \mathcal{N}( 0, \sigma^2_{y_{ }^{(1 \rightarrow 2)}} )$.

  \item[(c)]
  When $\frac{q_{}^{(1)}}{q_{}^{(2)}} \notin \mathbb{N} $,
  the relation \eqref{eq:y_relationB1} no longer holds.
  Hence, the distribution of  $y_{}^{(2)}$  should be obtained from $Y^{(2)}$ by using \eqref{eq:ypdf}.
\end{enumerate}
These cases complete the justification of Proposition \ref{pro:secondq}.

\section{Justification of Proposition  \ref{pro:secondr}  }
As the results are independent of the spatial index $m$, we drop the index for clarity.
We have the following two cases.
\begin{enumerate}
  \item[(a)]
  When $q_u^{(2)} \ge 2 ~\text{and}~ \frac{q_u^{(1)}}{q_u^{(2)}} \in \mathbb{N}, \forall u$,
  applying IDCT to  $ y_{ }^{(2)} $ and $ {y_{ }^{(1 \rightarrow 2)}}$ in \eqref{eq:jpeg_APR_relationA1},
  and using the relation \eqref{eq:Nx1} and \eqref{eq:Ny1}, we get
 \begin{equation}\label{eq:jpeg_APR_relationB1}
 \Pr \left \{  x_{ }^{(2)}   = - {x_{ }^{(1 \rightarrow 2)}} \right \} \approx 1
 \end{equation}
 Note that the rounding noise $| {x_{ }^{(1 \rightarrow 2)}} | \le 0.5$,
 which indicates that $\Pr \left \{  | x_{ }^{(2)}  | < 0.5 \right \} \approx 1$,
 and thus $\Pr \left \{  [ x_{ }^{(2)}  ] = 0 \right \} \approx 1$.
 Using the relation \eqref{eq:round_aux_relation}, we will have
  \begin{equation}\label{eq:jpeg_APR_relationC1}
 \Pr \left \{  x_{ }^{(2 \rightarrow 3)}   = - {x_{ }^{(2)}} \right \} \approx 1
 \end{equation}
 It implies that
 \begin{equation}\label{eq:jpeg_APR_relationD1}
 \Pr \left \{  x_{ }^{(2 \rightarrow 3)}   =  x_{ }^{(1 \rightarrow 2)}  \right \} \approx 1
 \end{equation}
As a result, we arrive $   x_{ }^{(2 \rightarrow 3)} \sim f_{  x_{ }^{(1 \rightarrow 2)} } $.

  \item[(b)]
  When the divisible quantization condition is not satisfied,  the relation \eqref{eq:jpeg_APR_relationB1} does not hold as well. In this case, $x_{ }^{(2)}$  follows Gaussian distribution.
  Having the relation \eqref{eq:round_aux_relation}, we arrive
  $   x_{ }^{(2 \rightarrow 3)} \sim \mathcal{Q^N}( \sigma^2_{x_{}^{(2)}}   ,1 ) $.
\end{enumerate}
These two cases complete the justification of Proposition \ref{pro:secondr}.

\section{Justification of Proposition  \ref{pro:kcycleq} } \label{appendixC}
The justification of Proposition  \ref{pro:kcycleq} is quite similar to that of Proposition  \ref{pro:secondq}.
We have the following lemma which facilitates the justification.
\newtheorem{lemma}[]{Lemma}
\begin{lemma} \label{pro:lemma}
The variance of a uniform distribution is the upper bound for the variance  of a quantized-Gaussian distribution when these two distributions have identical support.

\vspace{0.3em}
\noindent Proof:
According to \cite{Sripad1977}, the distribution of the quantization noise of a signal $Z$ with quantization step $q$ can be expressed as
\begin{equation}\label{eq:pdf_q_noise}
\begin{split}
f_{z}(s)
    &= \frac{1}{q}  + \frac{1}{q} \sum_{n \in \mathbb{Z}, n \neq 0} \Psi_{Z}(\frac{2 \pi n}{q}) e^{ -j \frac{ 2 \pi n }{q} s}   ,
    s \in [-\frac{q}{2}, \frac{q}{2} )
 \end{split}
\end{equation}
where $\Psi_{Z}(\cdot)$ is the characteristic function of $Z$.
When $Z \sim \mathcal{N}(0, \sigma^2)$, we have
\begin{equation}\label{eq:pdf_gau_q_noise}
\begin{split}
f_{z}(s)
    &= \frac{1}{q} + \frac{2}{q} \sum_{n=1}^{\infty}  e^{-\frac{ 2\pi^2 n^2 \sigma^2 }{q^2}} \cos( \frac{2\pi n }{q} s)     ,~ s \in [-\frac{q}{2}, \frac{q}{2} )
\end{split}
\end{equation}
Its variance is given by
\begin{equation}\label{eq:var_gau_q_noise}
    \sigma^2_{z}  = \frac{q^2}{12}  + \frac{q^2}{\pi^2} \sum_{n=1}^{\infty} \frac{(-1)^n}{n^2}  e^{-\frac{ 2\pi^2 n^2 \sigma^2 }{q^2}}
\end{equation}
Note that the first term in \eqref{eq:var_gau_q_noise} corresponds to the variance of a uniform distribution with support $[-\frac{q}{2}, \frac{q}{2} )$,
and the second term is a convergent alternating series.
As a result, $\sigma^2_{z}  \le \frac{q^2}{12} $,
and this completes the proof of Lemma 1.
\end{lemma}

We have the following three cases for $k \ge 2$.
\begin{enumerate}
  \item[(a)]
When $ {q}_{}^{(k+1)} = {1}$,
similar to \eqref{eq:y_relationD1}, we have
\begin{equation}\label{eq:y_relationE}
{y}_{ }^{(k+1)}
= - \left(  {y}_{ }^{(k \rightarrow k+1)} -   [ { {y}_{ }^{(k \rightarrow k+1)}} ]    \right)
\end{equation}
Since $ {y}_{ }^{(k \rightarrow k+1)} \sim \mathcal{N}(0, \sigma^2_{{y}_{ }^{(k \rightarrow k+1)}})$, 
hence we arrive ${y}_{ }^{(k+1)}$ $\sim$ $\mathcal{Q^N}(\sigma^2_{{y}_{ }^{(k \rightarrow k+1)}},1)$.

  \item[(b)]
When ${q}_{}^{(k+1)} \ge 2 $ and $\frac{q_{ }^{(k)}}{q_{ }^{(k+1)}} \in \mathbb{N} $,
according to Proposition \ref{pro:secondr} and \ref{pro:kcycler}, we know that
${ x_{ }^{(k+1 \rightarrow k+2)}   }$  follows quantized-Gaussian distribution, where the quantization step is $1$.
Based on Lemma \ref{pro:lemma}, we know that
\begin{equation}\label{eq:variance_bound}
  \sigma_{ x_{ }^{(k \rightarrow k+1)}   }^{2} \le \frac{1}{12}
\end{equation}
Using the DCT relationship between $y_{ }^{(k \rightarrow k+1)}$ and $x_{ }^{(k \rightarrow k+1)}$,
we can obtain that
\begin{equation}\label{eq:variance_bound}
  \sigma_{ y_{ }^{(k+1 \rightarrow k+2)}   }^{2} \le
  \max_{m} \left( \sigma_{ x_{m}^{(k+1 \rightarrow k+2)}   }^{2} \right) \le \frac{1}{12}
\end{equation}
Similar to \eqref{eq:3sigma} and \eqref{eq:jpeg_APR_relationA1}, we have
 \begin{equation}\label{eq:3sigma2}
    \begin{split}
         \Pr \left\{  \left[ \frac{y_{ }^{(k \rightarrow k+1)}}{q_{ }^{(k+1)}} \right] = 0  \right\}
         \approx 1
     \end{split}
 \end{equation}
and
 \begin{equation}\label{eq:jpeg_APR_relationA2}
 \Pr \left \{  y_{ }^{(k+1)}   = - {y_{ }^{(k \rightarrow k+1)}} \right \} \approx 1
 \end{equation}
Since
$y_{ }^{(k \rightarrow k+1)} \sim  \mathcal{N}( 0, \sigma^2_{y_{ }^{(k \rightarrow k+1)}} )$,
we arrive $ y_{ }^{(k+1)} \sim  \mathcal{N}( 0, \sigma^2_{y_{ }^{(k \rightarrow k+1)}} )$.

  \item[(c)] $\frac{q_{}^{(k)}}{q_{}^{(k+1)}} \notin \mathbb{N} $:
  The relation \eqref{eq:y_relationB1} no longer holds.
  Hence, the distribution of  $y_{}^{(k+1)}$  should be obtained from $Y^{(k+1)}$ by using \eqref{eq:ypdf}.
\end{enumerate}
These cases complete the justification of Proposition \ref{pro:kcycleq}.

\section{Justification of Proposition  \ref{pro:kcycler} }
As the results are independent of the spatial index $m$, we drop the index for clarity.
For $k \ge 2$, we have the following two cases.
\begin{enumerate}
  \item[(a)]
  When $q_u^{(k+1)} \ge 2 ~\text{and}~ \frac{q_u^{(k)}}{q_u^{(k+1)}} \in \mathbb{N}, \forall u$,
    applying IDCT to  $ y_{ }^{(2)} $ and $ {y_{ }^{(1 \rightarrow 2)}}$ in \eqref{eq:jpeg_APR_relationA2},
  and using the relation \eqref{eq:Nx1} and \eqref{eq:Ny1}, we get
 \begin{equation}\label{eq:jpeg_APR_relationB}
 \Pr \left \{  x_{ }^{(k+1)}   = - {x_{ }^{(k \rightarrow k+1)}} \right \} \approx 1
 \end{equation}
 Note that the rounding noise $| {x_{ }^{(k \rightarrow k+1)}} | \le 0.5$,
 which indicates that $\Pr \left \{  | x_{ }^{(k+1)}  | < 0.5 \right \} \approx 1$,
 and thus $\Pr \left \{  [ x_{ }^{(k+1)}  ] = 0 \right \} \approx 1$.
 Using the relation \eqref{eq:round_aux_relation}, we will have
  \begin{equation}\label{eq:jpeg_APR_relationC}
 \Pr \left \{  x_{ }^{(k+1 \rightarrow k+2)}   = - {x_{ }^{(k)}} \right \} \approx 1
 \end{equation}
 This implies that
 \begin{equation}\label{eq:jpeg_APR_relationD}
 \Pr \left \{  x_{ }^{(k+1 \rightarrow k+2)}   =  x_{ }^{(k \rightarrow k+1)}  \right \} \approx 1
 \end{equation}
As a result, we arrive $   x_{ }^{(k+1 \rightarrow k+2)} \sim f_{  x_{ }^{(k \rightarrow k+1)} } $.

  \item[(b)]
  If the divisible quantization condition does not hold,  the relation \eqref{eq:jpeg_APR_relationB} does not hold as well. In this case, $x_{ }^{(k+1)}$  follows Gaussian distribution.
  Having the relation \eqref{eq:round_aux_relation}, we arrive
  $   x_{ }^{(k+1 \rightarrow k+2)} \sim \mathcal{Q^N}( \sigma^2_{x_{}^{(k+1)}}   ,1 ) $.
\end{enumerate}
These two cases complete the justification of Proposition \ref{pro:kcycler}.

\section{Justification of Proposition \ref{pro:minimum}}

\subsection{Justification of $   Var \big[ y_{}^{(2)}({  \dot{q}  }) \big]   \le
            Var \big[ y_{}^{(2)}({  q^*  }) \big]  $}
Let $\dot{q} = q$ and $q^* = nq$ $(q, n \in \mathbb{N})$.

When $n=1$, we have $ \dot{q} = q^*$.  
The relation $Var \big[ y_{}^{(2)}({  \dot{q}  }) \big]   =  Var \big[ y_{}^{(2)}({  q^*  }) \big]$
is universally satisfied in this case.
Hence in the following we only need to discuss the case when $n \ge 2$, which implies that  $q^* = nq \ge 2$.

From \eqref{eq:y_relationC1} we can derive
\begin{equation}\label{eq:y_relationC}
\begin{split}
{y}_{ }^{(2)}
= -\left (  {y}_{ }^{(1 \rightarrow 2)} -  \left [ \frac{ {y}_{ }^{(1 \rightarrow 2)}}{q_{ }^{(2)}} \right] {q_{ }^{(2)}}  \right).
\end{split}
\end{equation}

Note that
the auxiliary noise ${y}_{ }^{(1 \rightarrow 2)}$ follows Gaussian distribution with $\sigma^2_{{y}_{ }^{(1 \rightarrow 2)}} \le 0.0833$.
Let ${y}_{ }^{(1 \rightarrow 2)} = -t$, we have
\begin{equation}\label{eq:app_prob_vartheta}
\begin{split}
    \Pr\{|t|< \frac{nq}{2} \} &\ge
    \Pr\{|t|<1\} \\
    &> \Pr\{| t |< 3 \sigma_{ {y}_{ }^{(1 \rightarrow 2)} }\}  \\
    & \approx 1.
\end{split}
\end{equation}
As a result, we can always assume $|t| < \frac{nq}{2} $, which implies $[ t / nq ] = 0$.
Then we can obtain
\begin{equation}\label{eq:app_t_relation}
\begin{split}
  \left(t - \left[ \frac {t} {q} \right] q \right)^2 \le \left(t - \left[ \frac {t} {nq} \right] nq \right)^2.
\end{split}
\end{equation}
The ``$=$'' holds when $|t| < \frac{q}{2}$, and the ``$<$" holds when $\frac{q}{2} \le |t| < \frac{nq}{2}$.

The variance of $y^{(2)}$ can be computed according to the relation indicated in \eqref{eq:y_relationC} and the probability distribution function (PDF) of ${y}_{ }^{(1 \rightarrow 2)}$, denoted by
$f_{{y}_{ }^{(1 \rightarrow 2)}}(t)$.
We have
\begin{equation*}
   Var \big[ y_{}^{(2)}({  \dot{q}  }) \big]  = \int_{-\infty}^{\infty} {\left( t - \left[ \frac {t} {q} \right] q \right)^2} f_{{y}_{ }^{(1 \rightarrow 2)}}(t)dt,
\end{equation*}
and
\begin{equation*}
   Var \big[ y_{}^{(2)}({  q^*  }) \big]  = \int_{-\infty}^{\infty} {\left( t - \left[ \frac {t} {nq} \right] nq \right)^2} f_{{y}_{ }^{(1 \rightarrow 2)}}(t)dt.
\end{equation*}
With the relation in \eqref{eq:app_t_relation}, as a result, we have
$   Var \big[ y_{}^{(2)}({  \dot{q}  }) \big]   \le    Var \big[ y_{}^{(2)}({  q^*  }) \big] $.

\subsection{Justification of $  Var \big[ y_{}^{(2)}({  q^*  }) \big] <   Var \big[ y_{}^{(2)}({  \ddot{q}  }) \big] $}
Let $q^* = q_\alpha$ and $\ddot{q} = q_\gamma$.
Let ${y}_{ }^{(1 \rightarrow 2)} = -t$ and we can always assume $|t|<1$ due to \eqref{eq:app_prob_vartheta}.

Suppose $\widetilde{Y}_{}^{(1)} = r q_\alpha$ ($r \in \mathbb{N}$), we have ${Y}_{}^{(2)} = r q_\alpha + t$.
Assume $r q_\alpha = s q_\gamma + p$, where $s \in \mathbb{Z}$ and  $p \in \{0, 1, \cdots, q_\gamma - 1\}$.

Define $d_1(t) = | y_{}^{(2)}({ q_\alpha }) |$ and $d_2(t) = | y_{}^{(2)}({ q_\gamma }) |$.
According to \eqref{eq:y_relationC}, we have
\begin{equation}\label{eq:app_d1}
  d_1(t) =  \left| t - \left[  \frac{t}{q_\alpha} \right] q_\alpha  \right|, \\  
\end{equation}
and
\begin{equation}\label{eq:app_d2}
  d_2(t) = \left| p+t - \left[  \frac{p+t}{q_\gamma} \right] q_\gamma  \right|.
\end{equation}

\subsubsection{ Case for $q_\alpha = 1$}

    When $q_\gamma = 2$,
    if all non-zero coefficients of $\tilde{Y}_{u}^{(1)}$ are even numbers, which is a rare case,
    we have $   Var \big( y_{}^{(2)}({ 1  }) \big) = Var \big( y_{}^{(2)}({ 2 }) \big) $.
    In common cases, there is at least one coefficient in $\tilde{Y}_{u}^{(1)}$ is an odd number,
    and we have $   Var \big( y_{}^{(2)}({ 1  }) \big) < Var \big( y_{}^{(2)}({ 2 }) \big) $.
    Thus we only discuss the situation when $q_\gamma \ge 3$.
    In this situation, we have $\left[   t/q_\gamma\right] = 0$ because of $|t|<1$.
    And therefore, we can reduce \eqref{eq:app_d1} and \eqref{eq:app_d2} into the following cases.
    \begin{equation}\label{eq:app_d1c1b}
        d_1(t) =  
        \begin{cases}
        1 - t \quad  &~~0.5 \le t < 1 \\
        t  \quad  &0 \le t < 0.5\\
        |t|  \quad  &-0.5 \le t < 0\\
        1-|t| \quad  &~-1 < t < -0.5
        \end{cases}
    \end{equation}
    \begin{equation}\label{eq:app_d2c1b}
        d_2(t) =
        \begin{cases}
        p + t \quad & 0 \le t < 1, ~ p+t < q_\gamma / 2      \\
        q_\gamma-p - t   & 0 \le t < 1, ~ p+t \ge q_\gamma / 2      \\
        |t|         & -1 < t<0, ~ p=0     \\
        p - |t|     & -1 < t<0, ~ p\ge1, p+t < q_\gamma / 2     \\
        q_\gamma-p + |t| & -1 < t<0, ~ p\ge1, p+t \ge q_\gamma / 2
        \end{cases}
    \end{equation}
    Note that because of the constraint $ 0 \le p \le q_\gamma - 1$ and $q_\gamma - p \ge 1$, we have the following relations.
    \begin{itemize}
      \item
      When $0.5 \le t < 1$ and $p+t < q_\gamma/2$, we have
      $ d_2(t) = p+t \ge 1-t = d_1(t)$,
      where ``=" holds when $p=0$ and $t=0.5$.

      \item
      When $0.5 \le t < 1$ and $p+t \ge q_\gamma/2$, we have
      $ d_2(t) = q_\gamma - p - t \ge 1-t = d_1(t)$,
      where ``=" holds when $p=q_\gamma - 1$.

      \item
      When $0 \le t < 0.5$ and $p+t < q_\gamma/2$, we have
      $ d_2(t) = p+t \ge t = d_1(t)$,
      where ``=" holds when $p=0$.

      \item
      When $0 \le t < 0.5$ and $p+t \ge q_\gamma/2$, we have
      $ d_2(t) = q_\gamma - p - t \ge 1-t > t = d_1(t)$.

      \item
      When $ -0.5 \le t < 0 $ and $p=0$, we have
      $ d_2(t) = |t| = d_1(t)$.

      \item
      When $ -1 < t < -0.5 $ and $p=0$, we have
      $ d_2(t) = |t| > 1 - |t| = d_1(t)$.

      \item
      When $ -0.5 \le t < 0 $, $p\ge 1$, and $p+t < q_\gamma/2$, we have
      $ d_2(t) = p - |t| \ge 1- |t| \ge |t| = d_1(t)$,
      where ``=" holds when $p=1$ and $t = -0.5$.

      \item
      When $ -0.5 \le t < 0 $, $p\ge 1$, and $p+t \ge q_\gamma/2$, we have
      $ d_2(t) = q_\gamma - p + |t| \ge 1 + |t|  > |t| = d_1(t)$.

      \item
      When $ -1 < t < -0.5 $ and $p\ge 1$, and $p+t < q_\gamma/2$, we have
      $ d_2(t) = p - |t| \ge 1 - |t| = d_1(t)$,
      where ``=" holds when $p=1$.

      \item
      When $ -1 < t < -0.5 $ and $p\ge 1$, and $p+t \ge q_\gamma/2$, we have
      $ d_2(t) = q_\gamma - p + |t| \ge 1 + |t| > 1 - |t| = d_1(t)$.

    \end{itemize}

    Based on the above relations, we can conclude that under most conditions, we get $d_1(t) < d_2(t)$, and only under rare conditions, we get $d_1(t) = d_2(t)$.
    As there are sufficient many non-zero dequantized DCT coefficients, we would have $   Var \big( y_{}^{(2)}({  q_\alpha  }) \big) < Var \big( y_{}^{(2)}({  q_\gamma  }) \big) $.

\subsubsection{ Case for $q_\alpha \ge 2$}

  Since $1$ is a divisor of any integer number, we have $q_\gamma \neq 1$.
  In other words, $q_\gamma \ge 2$.
  In this case,  $\left[   t/q_\alpha  \right] = 0$ because of $|t|<1$, and thus \eqref{eq:app_d1} reduces to
    \begin{equation}\label{eq:app_d1c2}
        d_1(t) =  
        \begin{cases}
          t  \quad  &0 \le t < 1\\
        |t|  \quad  &-1 < t < 0
        \end{cases}
    \end{equation}
  and \eqref{eq:app_d2} still reduces to \eqref{eq:app_d2c1b}.
  We have the follow relations.

  \begin{itemize}
      \item
      When $0 \le t < 1$ and $p+t < q_\gamma/2$, we have
      $ d_2(t) = p+t \ge t = d_1(t)$,
      where ``=" holds when $p=0$.

      \item
      When $0 \le t < 1$ and $p+t \ge q_\gamma/2$, we have
      $ d_2(t) = q_\gamma - p + t \ge 1 + t > t = d_1(t)$.

      \item
      When $-1 < t < 0$ and $p=0$, we have
      $ d_2(t) = |t| = d_1(t)$.

      \item
      When $-0.5 \le t < 0$, $p = 1$, and $p+t < q_\gamma/2$, we have
      $ d_2(t) = 1 -|t| \ge |t| = d_1(t)$,
      where ``=" holds when $t=-0.5$.

      \item
      When $-1 < t < -0.5 $, $p = 1$, and $p+t < q_\gamma/2$, we have
      $ d_2(t) = 1 -|t| < |t| = d_1(t)$.

      \item
      When $-1 < t < 0$, $p \ge 2$, and $p+t < q_\gamma/2$, we have
      $ d_2(t) = p - |t| \ge 2 -|t| > |t| = d_1(t)$.

      \item
      When $-1 < t < 0$, $p \ge 1$, and $p+t \ge q_\gamma/2$, we have
      $ d_2(t) = q_\gamma - p + |t| \ge 1+|t| >|t| = d_1(t)$.

  \end{itemize}

  Based on the above relations, we can conclude that under most conditions, we have $d_1(t) < d_2(t)$, and under rare conditions, we have $d_1(t) \ge d_2(t)$.
  As there are sufficient many non-zero dequantized DCT coefficients, we would have $   Var \big( y_{}^{(2)}({  q_\alpha  }) \big) < Var \big( y_{}^{(2)}({  q_\gamma  }) \big) $.

  Based on the above two cases, we can conclude that $   VAR \big( y_{}^{(2)}({  q^* }) \big) < VAR \big( y_{}^{(2)}({  \ddot{q}  }) \big) $.

\section{Justification of Proposition \ref{pro:id2}}\label{sec:appendix}

  We already know that:
  \begin{itemize}
      \item
      The spatial auxiliary noise is the IDCT transform of quantization noise (refer to \eqref{eq:Nx1}). We rewrite it as:
      \begin{equation}\label{appeq:RelationA}
           x_{m}^{(k)} = \sum_{u=1}^{64}
               \tilde{c}_{m,u} y_{u}^{(k)} , \forall m,
      \end{equation}
      where $\tilde{c}_{m,u}$ is the weight for the inverse transform, and
      $\sum_{m=1}^{64} (\tilde{c}_{m,u})^2 = 1$.

    \item
    The DCT auxiliary noise is the DCT transform of rounding noise  (refer to \eqref{eq:Ny1}). We rewrite it as:
      \begin{equation}\label{appeq:RelationB}
           y_{u}^{(k \rightarrow k+1)} = \sum_{m=1}^{64}
              c_{u,m} x_{m}^{(k \rightarrow k+1)} , \forall u,
      \end{equation}
      where $c_{u,m}$ is the weight for the forward transform, and
      $\sum_{m=1}^{64} (c_{u,m})^2 = 1$.
%

  \end{itemize}


  The variance of a signal is no smaller than that of its quantization noise.
  Hence with \eqref{eq:round_aux_relation}, we know that
  \begin{equation}\label{appeq:sigmaA}
    	 \sigma^2_{x_m^{(2 \rightarrow 3)}}  \leq   \sigma^2_{x_m^{(2)}}, ~~\forall m.
  \end{equation}

  When $q_{u}^{(k+1)}=1$,
  applying \eqref{appeq:RelationA}, \eqref{eq:y_relationD1}, and \eqref{appeq:RelationB} in sequence,
  we obtain
  \begin{equation}\label{appeq:signal}
    \begin{split}
    	 x_m^{(2)} &= \sum_{u=1}^{64}   \tilde{c}_{m,u} {y_u^{(2)}} \\
    & = \sum_{u=1}^{64}   \tilde{c}_{m,u}
    \left(   - y_u^{(1 \rightarrow 2)} + [ y_u^{(1 \rightarrow 2) }]   \right) \\
    & = \sum_{u=1}^{64}   \tilde{c}_{m,u}
    \left( -\sum_{m=1}^{64} c_{u,m} x_{m}^{(1 \rightarrow 2)}
    + \left[ \sum_{m=1}^{64} c_{u,m} x_{m}^{(1 \rightarrow 2)} \right]
    \right) \\
     & = -x_{m}^{(1 \rightarrow 2)}
    + \sum_{u=1}^{64} \tilde{c}_{m,u}
    \left[ \sum_{m=1}^{64} c_{u,m} x_{m}^{(1 \rightarrow 2)} \right] \\
    \end{split}
  \end{equation}
  The first term of the last equation in \eqref{appeq:signal} is derived because DCT is a kind of unitary transform.

    Let
      \begin{equation}\label{appeq:omega}
     \omega_{m} \triangleq     \sum_{u=1}^{64} \tilde{c}_{m,u}
     \left[ \sum_{m=1}^{64} c_{u,m} x_{m}^{(1 \rightarrow 2)} \right].
      \end{equation}
  Assume $x_{m}^{(1 \rightarrow 2)}$ and $\omega_{m}$ are independent.
  As a result,
  \begin{equation}\label{appeq:sigmaB}
    	 \sigma^2_{x_m^{(2 )}}  =  \sigma^2_{x_m^{(1 \rightarrow 2)}} + \sigma^2_{\omega_{m}} \leq \sigma^2_{x_m^{(1 \rightarrow 2)}}, ~~\forall m.
  \end{equation}

  Apply the relations in \eqref{appeq:sigmaA} and  \eqref{appeq:sigmaB}, we arrive
  \begin{equation}\label{appeq:sigmaXmD}
    	     \sigma^2_{x_m^{(2 \rightarrow 3)}} \leq \sigma^2_{x_m^{(1\rightarrow 2)}}, ~~\forall  m.
  \end{equation}
\ifCLASSOPTIONcaptionsoff
  \newpage
\fi


\bibliographystyle{IEEEtran}
\bibliography{IEEEabrv,noise_model}

\end{document}